\documentclass[preprint,review,12pt, authoryear]{elsarticle}

\usepackage{lineno,hyperref}
\usepackage{tikz}
\usepackage{amssymb}
\usepackage{amsmath}
\usepackage{amsfonts} 
\usepackage{relsize}
\usepackage{url}
\usepackage{svg}
\usepackage{subfigure}
\usepackage{graphicx}
\usepackage{multirow}
\usepackage{booktabs}
\usepackage{mathtools}

\journal{Expert Systems with Applications}

\begin{document}

\begin{frontmatter}

\title{Bayesian implementation of Targeted Maximum Likelihood Estimation for uncertainty quantification in causal effect estimation}

\author{Saideep Nannapaneni\corref{cor1}}
\ead{saideepn@apporchid.com}
\cortext[cor1]{Corresponding author}
\author{Joseph Sakaya}
\ead{josephs@apporchid.com}
\author{Kyle Caron}
\ead{kylec@apporchid.com}
\author{Pedro HM Albuquerque}
\ead{pedroa@apporchid.com}
\author{Zaid Tashman}
\ead{zaidt@apporchid.com}

\address{App Orchid Inc.\\
6111 Bollinger Canyon Rd, Suite 570,
San Ramon, CA, 94583, USA\\[-8mm]}

\begin{abstract}

Robust decision making involves making decisions in the presence of uncertainty and is often used in critical domains such as healthcare, supply chains, and finance. Causality plays a crucial role in decision-making as it predicts the change in an outcome (usually a key performance indicator) due to a treatment (also called an intervention). To facilitate robust decision making using causality, this paper proposes three Bayesian approaches of the popular Targeted Maximum Likelihood Estimation (TMLE) algorithm, a flexible semi-parametric double robust estimator, for a probabilistic quantification of uncertainty in causal effects with binary treatment, and binary and continuous outcomes. In the first two approaches, the three TMLE models (outcome, treatment, and fluctuation) are trained sequentially. Since Bayesian implementation of treatment and outcome yields probabilistic predictions, the first approach uses mean predictions, while the second approach uses both the mean and standard deviation of predictions for training the fluctuation model (targeting step). The third approach trains all three models simultaneously through a Bayesian network (called BN-TMLE in this paper). The proposed approaches were demonstrated for two examples with binary and continuous outcomes and validated against classical implementations. This paper also investigated the effect of data sizes and model misspecifications on causal effect estimation using the BN-TMLE approach. Results showed that the proposed BN-TMLE outperformed classical implementations in small data regimes and performed similarly in large data regimes.
\end{abstract}



\end{frontmatter}


\section{Motivation and Introduction}
\label{sec:intro}

Causality is a fundamental concept widely used in various domains of science, business, and engineering, and is described as a notion that two variables have a cause-and-effect relationship between them, i.e., the cause is at least partly responsible for the effect \citep{yao2021survey}. Causal inference refers to the process of determination of a presence of causal relationship between two variables and causal effect estimation is defined as the numerical quantification of the causal effect between the two variables.

Causal inference and effect estimation, together referred to as causal analysis in this paper, are useful for knowledge discovery, i.e., discovering previously unknown cause-effect relationships between variables, and this new knowledge may later be used in decision making when the ``cause" variable is controllable. Causal analysis has previously been used in various domains for actionable decision making such as healthcare decision making \citep{prosperi2020causal, sanchez2022causal}, marketing \citep{varian2016causal, hair2021data}, public policy \citep{glass2013causal, gangl2010causal, athey2015machine}, education \citep{kaur2019causal, cordero2018causal}, engineering design \citep{naser2022causality, siebert2023applications}, process improvement \citep{chen2021ontology, meng2020localizing}, and operations management \citep{ho2017om, mithas2022causality}. 

Bayesian methods are often employed in decision making due to their ability to systematically quantify and aggregate uncertainty in data and models, and provide probabilistic estimation and uncertainty quantification of output of interest, which can be used for risk-informed decision making \citep{trimmer2011decision, kochenderfer2015decision}. In addition, the Bayesian paradigm enables the fusion of data from multiple sources (e.g., prior / expert knowledge and observational data) \citep{dai2023bayesian} for comprehensive modeling and uncertainty quantification. Moreover, probabilistic predictions from Bayesian models enable a direct interpretation of the variable to lie within a certain interval with a certain probability (computed from the probabilistic prediction), whereas such an interpretation is not possible with traditional confidence intervals \citep{hespanhol2019understanding}. 

Probabilistic programming is a novel programming paradigm that provides tools and techniques for probabilistic modeling, reasoning, inference, and prediction following the Bayesian modeling framework \citep{gordon2014probabilistic, tran2017deep}. It facilitates flexibility and modularity, enabling quick modification of distribution types and parameters of variables, model formulation, and retraining models effectively. It also provides abstractions to represent and train complex models (such as hierarchical models) efficiently. Some widely used probabilistic programming languages include Stan \citep{carpenter2017stan}, Pyro \citep{bingham2019pyro}, PyMC \citep{patil2010pymc}, and NumPyro \citep{phan2019composable}. 

Due to the benefits of Bayesian modeling framework and availability of the probabilistic programming paradigms, this paper uses these ideas to illustrate (Bayesian) probabilistic prediction and uncertainty quantification in causal effect estimation for it can be used later for robust decision making considering uncertainty.

Two of the most popular approaches to infer causality are Randomized Controlled Trials (RCTs) and Inference from Observational Data. In RCTs, the researchers have full control over the design and implementation of the data generation process to study causality. Observational data is defined as naturally-occurring data that is collected without any control or intervention into the system. In observational studies, the confounders are not controlled for in the data generation process; therefore, the available data needs to be evaluated carefully to remove any bias due to confounders when inferring causality. 

This paper primarily focuses on observational studies, where causality needs to be inferred between two variables given data on them in the presence of confounders. However, it should be noted that the methods proposed for observational data can also be implemented for RCTs. The Potential Outcomes Framework (POF), also referred to as Counterfactual Framework or Neyman–Rubin Potential Outcomes or Rubin Causal Model, a widely used framework for causal inference from observational data, is based on counterfactual predictions, i.e., the outcome prediction for the unobserved scenarios \citep{rubin2005causal}. 

Several techniques exist to estimate causal effects based on POF such as Propensity Score-based methods \citep{shiba2021using, shi2022learning}, Meta Learners \citep{kunzel2019metalearners, acharki2023comparison}, Double/Debiased Machine Learning (DML) \citep{chernozhukov2018double}, and Doubly Robust Estimation (DRE) \citep{funk2011doubly, bang2005doubly}. Propensity score-based methods calculate causal effects using propensity scores, which are computed using simple models such as logistic regression, and these propensity scores are used to adjust the influence of confounders. Two widely propensity-based methods are Propensity score matching (PSM) \citep{li2013using} and Inverse Propensity of Treatment Weighting (IPTW) \citep{shiba2021using}. 

Meta learners use a variety of predictive models to estimate outcomes with and without treatments, and use these estimated outcomes to compute causal effects. Different types of Meta Learners exist in literature such as T-Learner, S-Learner, X-Learner, and DR-Learner \citep{chernozhukov2024applied}. Double/Debiased Machine Learning (DML) and Doubly Robust Estimation (DRE) are two novel approaches to obtain unbiased estimates of causal effects. Both DML and DRE train multiple models to predict the outcome and propensity scores but use these predictions differently to obtain unbiased causal estimates. DML uses the principle of orthogonality to remove bias in causal estimation. DRE is doubly robust as it enables effective causal effect estimation when either the outcome or propensity model is incorrectly specified. Two popular DRE techniques are the Augmented Inverse Probability Weighting (AIPW) \citep{kurz2022augmented} and Targeted Maximum Likelihood Estimation (TMLE) \citep{van2011targeted}.

TMLE is a flexible semi-parametric DRE technique that uses the principles of targeted learning \citep{van2006targeted} for unbiased causal effect estimation. The semi-parametric property refers to the use of nonparametric data-driven models (both individual and ensemble models) for outcome and propensity score predictions, and the use of parametric approaches to implement principles of targeted learning to optimize the bias-variance trade-off by adjusting the initial outcome predictions with propensity scores leading to unbiased causal estimates and lowest possible asymptotic variance, which can be used to obtain confidence intervals of the causal estimate \citep{smith2023application, luque2018targeted}. Since TMLE combines the benefits of machine learning used for outcome and propensity model training with statistical rigor in obtaining statistical confidence intervals of the causal estimate, it has become a widely popular technique, and therefore, it is considered in this paper. 

The goal of this paper is to combine the algorithmic benefits of TMLE (such as semi-parametric efficiency) with the modeling flexibility offered by Bayesian modeling paradigm to implement a Bayesian TMLE algorithm and illustrate probabilistic uncertainty quantification in causal effect estimation for robust decision making using the probabilistic programming paradigm. This paper presents three formulations for Bayesian implementation of the TMLE algorithm. Two formulations train various models trained in TMLE framework sequentially, and the third formulation integrates all the models through a Bayesian network and trains them simultaneously. 

\textbf{Paper Contributions:} The overall contributions made through this paper are: (1) Three Bayesian formulations of the Targeted Maximum Likelihood Estimation (TMLE) algorithm, (2) A sampling-based approach for probabilistic prediction and uncertainty quantification  of causal effect estimation, (3) Two simulation studies illustrating the proposed approaches for both binary and continuous outcomes, and their comparison against classical TMLE implementations and underlying known true causal effects, and (4) Investigation of the effect of data size and model misspecification on the accuracy of proposed Bayesian TMLE against classical implementation and known true causal effect.

\textbf{Paper Organization:} The remainder of the paper is organized as follows. Section \ref{sec:back} reviews the Targeted Maximum Likelihood Estimation (TMLE) algorithm for causal estimation and Bayesian modeling paradigm. Section \ref{sec:uq} details the three Bayesian TMLE formulations for a binary treatment, and with binary and continuous outcomes. Section \ref{sec:exp} details two case studies describing the application of the proposed approaches for uncertainty quantification in causal estimate (ATE) and compares the results with classical TMLE analysis. It also investigates the effect of data size and model misspecification on ATE estimation.  Section \ref{sec:conc} provides concluding remarks and directions for future work.

\section{Background}
\label{sec:back}

\subsection{Targeted Maximum Likelihood Estimation (TMLE)}
\label{subsec:tmle}
Three models are trained as part of the TMLE framework: an outcome model, a propensity model, and a fluctuation model. The outcome model predicts the outcome ($Y$) given the treatment variable ($A$) and a set of confounders ($\mathbf{X}$), the propensity model predicts the treatment variable ($A$) given a set of confounders ($\mathbf{X}$), and fluctuation model adjusts the outcome predictions to obtain an unbiased estimate of causal effect. 
We briefly review the training of the three models and their usage for estimating measures of causal inference.

We can choose any machine learning model formulation (e.g. generalized linear models, polynomial models, neural networks, decision trees, Gaussian process models, or even ensemble formulations) that provides a good fit for the outcome and propensity models. The ensemble modeling approach is related to the idea of a Super Learner \citep{van2007super}, which uses a cross validation approach to assess performance of multiple machine learning models and create an optimal combination of these models (weighted average) for the purpose of improving the modeling performance. The outcome and propensity models are given below in Equations \ref{eqn:outcome} and \ref{eqn:propensity} respectively for a binary outcome.

\begin{equation}
\label{eqn:outcome}
\begin{split}
     \mbox{logit}(Y) &= g(\mathbf{X}, A; \boldsymbol{\theta_Y})\\
     P(Y=1|\mathbf{X}, A) &= \mbox{expit}(g(\mathbf{X}, A; \boldsymbol{\theta_Y}))
\end{split}
\end{equation}

\begin{equation}
\label{eqn:propensity}
\begin{split}
    \mbox{logit}(A) &= h(\mathbf{X}; \boldsymbol{\theta_A})\\
    P(A=1|\mathbf{X}) &= \mbox{expit}(h(\mathbf{X}; \boldsymbol{\theta_A}))
\end{split}
\end{equation}

In Equation  \ref{eqn:outcome}, $g(\mathbf{X}, A; \boldsymbol{\theta_Y})$ is any machine learning formulation used to model the outcome $Y$, and $\boldsymbol{\theta_Y}$ represent the model parameters.  Similarly, in Equation \ref{eqn:propensity}, $h(\mathbf{X}; \boldsymbol{\theta_A})$ is any formulation used to model the treatment $A$, and $\boldsymbol{\theta_A}$ are the relevant model parameters.

The propensity model is used to estimate a ``clever covariate" value corresponding to each data record, and these clever covariate values are later used for training the fluctuation model. The clever covariate $H(A, \mathbf{X})$ is defined as shown in Equation \ref{eqn:clever} \citep{schuler2017targeted}. 

\begin{equation}
\label{eqn:clever}
    H(A, X) = \dfrac{I(A=1)}{P(A=1|\mathbf{X})} - \dfrac{I(A=0)}{P(A=0|\mathbf{X})}
\end{equation}

In Equation \ref{eqn:clever}, $I(A=a), a \in \{ 0,1\}$ is the indicator function. For example, $I(A=1) = 1$ when $A=1$ and $0$ when $A=0$. $P(A=1|\mathbf{X})$ and $P(A=0|\mathbf{X})$ are the probabilities of being in the treatment and control groups respectively, given the confounder values.

The third model combines the predictions from the outcome and propensity models, and provides updated outcome predictions to obtain unbiased causal effect estimates. 
The fluctuation model formulation is described in Equation \ref{eqn:fluctuation} \citep{schuler2017targeted}.

\begin{equation}
\label{eqn:fluctuation}
    \mbox{logit}(Y) = \mbox{logit}(Y_{init}) + \epsilon H(A, \mathbf{X})
\end{equation}

In Equation \ref{eqn:fluctuation}, $Y_{init}$, $H(A, \mathbf{X})$, and $\epsilon$ are the initial outcome prediction, clever covariate, and the fluctuation parameter respectively. The initial outcome prediction, $Y_{init}$ can be calculated as $\mathbb{E}(Y|A, \mathbf{X})= \mbox{expit}(g(\mathbf{X}, A; \boldsymbol{\theta_Y}))$, where $\mathbb{E}$ is the expectation function.

Even though any machine learning formulation can be used for training outcome and propensity models, generalized linear models (GLM) such as logistic regression are typically used for training the fluctuation model \citep{van2011targeted}. In addition to the above formulation of the clever covariate and fluctuation model, there exists an alternative formulation with two clever covariate terms corresponding to control and treatment groups of data, defined below in Equation \ref{eqn:clever2}, and the corresponding fluctuation model formulation is described in Equation \ref{eqn:fluctuation2} \citep{frank2024implementing}.

\begin{equation} \label{eqn:clever2}
\begin{split}
H_1(A, \mathbf{X}) = \dfrac{I(A=1)}{P(A=1|\mathbf{X})}\\
H_0(A, \mathbf{X}) = \dfrac{I(A=0)}{P(A=0|\mathbf{X})}\\
\end{split}
\end{equation}

\begin{equation}
\label{eqn:fluctuation2}
    \mbox{logit}(Y) = \mbox{logit}(Y_{init}) + \sum_{a = 0, 1} \epsilon_a H_a(A, \mathbf{X})
\end{equation}

In Equation \ref{eqn:clever2}, 
$\epsilon_a$ represents the fluctuation parameter for $A=a, a \in \{ 0,1\}$. In this formulation, there exist two fluctuation parameter terms $\epsilon_0$ and $\epsilon_1$, which control the amount of fluctuation of the initial outcome in control ($A=0$) and treatment ($A=1$) respectively.  In this paper, we refer to Equations \ref{eqn:fluctuation} and \ref{eqn:fluctuation2} as one-parameter and two-parameter fluctuation models. We primarily use the one-parameter fluctuation model formulation to illustrate the proposed Bayesian TMLE approaches; however, the one-parameter formulation can easily be swapped with the two-parameter formulation. 


Following the potential outcomes framework for causal inference, we obtain model predictions with and without treatment using all the three models denoted as $Y_f(1, \mathbf{X})$ and $Y_f(0, \mathbf{X})$, and calculate the Average treatment Effect (ATE) as 

\begin{equation}
\label{eqn:ate}
\mbox{ATE} =  \mathop{{}\mathbb{E}}[Y_f(1, \mathbf{X})] - \mathop{{}\mathbb{E}}[Y_f(0, \mathbf{X})] = \mathop{{}\mathbb{E}}[Y_f(1, \mathbf{X}) - Y_f(0, \mathbf{X}))]
\end{equation}






More details regarding the TMLE framework are available in \citet{luque2018targeted} and \citet{schuler2017targeted}. 

\subsection{Bayesian modeling paradigm}
\label{subsec:bayes}
Bayesian modeling paradigm corresponds to the use of Bayes theorem for model training (model parameter estimation), often referred to as Bayesian inferenence and model prediction. We detail below the application of Bayesian paradigm for classification analysis. Let $D = {\mathbf{X}_i, Y_i}, i=1\dots n$ represent a dataset on a set of input features $\mathbf{X}$ and a continuous output $Y$. Consider the model formulation in Equation \ref{eqn:model}, where $\boldsymbol{\theta}$ corresponds to the model parameters, $\epsilon$ is the measurement error corresponding to the measurement in $Y$, and $h(.)$ represents the functional relationship between inputs and output.

\begin{equation}
\label{eqn:model}
    Y = h(\mathbf{X};\boldsymbol{\theta} + \xi
\end{equation}

The measurement error is typically modeled using a symmetric distribution with a mean value of 0, to account for both positive and negative values of the measurement error. A Gaussian distribution is commonly used to model the measurement error, $\xi \sim N(0, \sigma_\xi)$. If $\sigma_\xi$ is unknown prior to model training, then it is also considered a learning parameter along with $\boldsymbol\theta$. Using the Bayes theorem, the model parameters can be estimated as shown in Equation \ref{eqn:bayes}, where $p(\boldsymbol\theta, \sigma_\xi|D)$, $p(D|\boldsymbol\theta, \sigma_\xi)$, $p(\boldsymbol\theta, \sigma_\xi)$, and $p(D)$ correspond to the posterior, prior, likelihood, and evidence terms.

\begin{equation}
\label{eqn:bayes}
p(\boldsymbol\theta, \sigma_\xi|D) = \dfrac{p(D|\boldsymbol\theta, \sigma_\xi)p(\boldsymbol\theta, \sigma_\xi)}{p(D)}
\end{equation}

The prior distribution, $p(\boldsymbol\theta, \sigma_\xi)$, encodes any prior knowledge that may be available about the model parameters. The likelihood corresponds the probability of observing the available data, $D$, conditioned on the model parameters. The posterior distribution, $p(\boldsymbol\theta, \sigma_\xi|D)$, presents the updated knowledge after integrating the likelihood and prior distribution. The likelihood term, $p(D|\boldsymbol\theta, \sigma_\xi)$, can be written as 

\begin{equation}
\label{eqn:like}
    p(D|\boldsymbol\theta, \sigma_\xi) = \prod_{i=1}^n p(Y_i|\mathbf{X}_i,\boldsymbol\theta, \sigma_\xi) p(\mathbf{X}_i) 
\end{equation}

Combining, Equations \ref{eqn:bayes} and \ref{eqn:like}, we have 

\begin{equation}\label{eqn:bapprox}
\begin{split}
    p(\boldsymbol\theta, \sigma_\xi|D) & = \dfrac{\prod_{i=1}^n p(Y_i|\mathbf{X}_i,\boldsymbol\theta, \sigma_\xi) p(\mathbf{X}_i)p(\boldsymbol\theta, \sigma_\xi)}{\prod_{i=1}^n p(Y_i|\mathbf{X}_i) p(\mathbf{X}_i)} \\
    & = \dfrac{\prod_{i=1}^n p(Y_i|\mathbf{X}_i,\boldsymbol\theta, \sigma_\xi)p(\boldsymbol\theta, \sigma_\xi)}{\prod_{i=1}^n p(Y_i|\mathbf{X}_i)}  \\
\end{split}
\end{equation}

Several algorithms, both exact and approximate, are available for the estimation of model parameters. Exact algorithms such as variable elimination and conjugate prior method \citep{koller2009probabilistic}, even though available, do not work for all variable types (binary, categorical, continuous) and probability distributions. When the denominator term in Equation \ref{eqn:bapprox} is intractable to compute, we use approximate algorithms to estimate the posterior distributions of model parameters. Since the denominator term is independent of the model parameters, Equation \ref{eqn:bapprox} can be simplified to $p(\boldsymbol\theta, \sigma_\xi|D) \propto \prod_{i=1}^n p(Y_i|\mathbf{X}_i,\boldsymbol\theta, \sigma_\xi) p(\boldsymbol\theta, \sigma_\xi)$. Approximate inference algorithms such as Markov Chain Monte Carlo (MCMC) methods \citep{gamerman2006markov, gelman1995bayesian}, Approximate Bayesian Computation (ABC) \citep{beaumont2010approximate}, and Stochastic Variational Inference (SVI) methods \citep{hoffman2013stochastic} are widely employed to estimate posterior distributions of model parameters.

\section{Bayesian TMLE implementation}
\label{sec:uq}
In this section, we discuss the implementation of the TMLE algorithm in the Bayesian modeling paradigm for uncertainty quantification in causal estimates. In this paper, we consider a binary treatment and a binary and continuous outcomes with no restriction on the data types of confounding variables. As mentioned in Section \ref{sec:intro}, we propose three formulations for Bayesian implementation of the TMLE algorithm. The first two formulations consider sequential training of the three TMLE models: outcome, propensity, and fluctuation, while the third formulation considers combining all three models through a Bayesian network and train the three models simultaneously.

Section \ref{subsec:sbtmle} discusses the two formulations involving sequentially training of TMLE models. The procedure for training the outcome and propensity models is the same; however, they differ in training the fluctuation model.  Section \ref{subsec:bn-tmle} discusses Bayesian network approach for simultaneous training of all three models. 

\subsection{Sequential Bayesian TMLE}
\label{subsec:sbtmle}

\subsubsection{Propensity model}

As we consider a binary treatment, the Bayesian propensity model formulation for the propensity formulation in Equation \ref{eqn:propensity} is given as

\begin{equation}
\label{eqn:bintbino_prop}
\begin{split}
\boldsymbol{\theta_A} &\sim \mbox{MVN}(\boldsymbol\mu_p, \boldsymbol\Sigma_p)\\
\eta_i &=  \mbox{expit}(h(\mathbf{X}_i; \boldsymbol{\theta_A}))\\
A_i|\mathbf{X}_i, \boldsymbol{\theta_A} &\sim \mbox{Bernoulli}(\eta_i)\\
p(\boldsymbol{\theta_A}|\mathbf{X}, A) &\propto p(\boldsymbol{\theta_A}) \prod_{i=1}^d p(A_i|\mathbf{X}_i,\boldsymbol{\theta_A})
\end{split}
\end{equation}

In Equation \ref{eqn:bintbino_prop}, $p(\boldsymbol{\theta_A})$ represents the joint prior distribution over the model parameters, which is assumed to be a Multivariate Gaussian distribution with $\boldsymbol\mu_p$ and $\boldsymbol\Sigma_p$ as its parameters (mean vector and covariance matrix). Here, a Multivariate Gaussian is used for illustration and it can be replaced with any prior distribution as required. $\eta_i$ represents the mean estimate of the treatment probability conditioned on the model parameter at any input $\mathbf{X}_i$.  The likelihood value  $A = A_i$ at any input  $\mathbf{X}_i$ can be calculated using a Bernoulli distribution with a parameter value of $\eta_i$. The posterior distributions of the model parameters can then be calculated using their prior distributions and the overall likelihood across all available $d$ data points.

\subsubsection{Outcome model}

Since both binary and continuous outcomes are considered, we describe model formulations for each of these two cases separately below.

\underline{\textit{Binary outcome:}} Similar to the Bayesian formulation of the propensity model in Equation \ref{eqn:bintbino_prop}, the Bayesian formulation for the outcome model with a binary outcome given in Equation \ref{eqn:outcome} is given below in Equation \ref{eqn:bintbino_out}.

\begin{equation}
\label{eqn:bintbino_out}
\begin{split}
\boldsymbol{\theta_Y} &\sim \mbox{MVN}(\boldsymbol\mu_o, \boldsymbol\Sigma_o)\\
\zeta_i &=  \mbox{expit}(g(\mathbf{X}_i, A_i; \boldsymbol{\theta_Y})\\
Y_i|\mathbf{X}_i, A_i, \boldsymbol{\theta_Y} &\sim \mbox{Bernoulli}(\zeta_i)\\
p(\boldsymbol{\theta_Y}|Y, \mathbf{X}, A) &\propto p(\boldsymbol{\theta_Y}) \prod_{i=1}^d p(Y_i|\mathbf{X}_i, A_i, \boldsymbol{\theta_Y})\\
\end{split}
\end{equation}

The Bayesian outcome model formulation in Equation \ref{eqn:bintbino_out} begins with the prior distribution over the model parameters, where we assumed a Multivariate Gaussian with parameters $\boldsymbol\mu_o$ and $\boldsymbol\Sigma_o$ .  $\zeta_i$ represents the mean estimate of the success probability conditioned on the model parameters for any input $\mathbf{X}_i$ and $A_i$. The likelihood of the observed value $Y=Y_i$ can be calculated using a Bernoulli distribution with a parameter value of $\zeta_i$. The posterior distributions of the outcome model parameters can then be calculated using their prior distributions and the overall likelihood across all available data points.

\underline{\textit{Continuous outcome:}} The outcome modeled formulation in the presence of a continuous outcome and associated Bayesian formulation are provided below in Equations \ref{eqn:regress} and \ref{eqn:bintconto_out} respectively.

\begin{equation}
\label{eqn:regress}
\begin{split}
 Y = g(\mathbf{X}, A; \boldsymbol{\theta_Y})+ \xi
\end{split}
\end{equation}

\begin{align}
\begin{split}
\sigma_\xi &\sim \mbox{HalfNormal}(\sigma_o) \\
\xi|\sigma_\xi &\sim \mbox{Normal}(0, \sigma_\xi)\\
\boldsymbol{\theta_Y} &\sim \mbox{MVN}(\boldsymbol\mu_o, \boldsymbol\Sigma_o)\\
\zeta_i &=  g(\mathbf{X}_i, A_i; \boldsymbol{\theta_Y})\\
Y_i|\mathbf{X}_i, A_i, \boldsymbol{\theta_Y}, \sigma_\xi &\sim \mbox{Normal}(\zeta_i, \sigma_\xi)\\
p(\boldsymbol{\theta_Y}, \sigma_\xi|Y, \mathbf{X}, A) &\propto p(\boldsymbol{\theta_Y})p(\sigma_\xi) \prod_{i=1}^d p(Y_i|\mathbf{X}_i, A_i, \boldsymbol{\theta_Y}, \sigma_\xi)
\end{split}
\label{eqn:bintconto_out}
\end{align}

In Equation \ref{eqn:regress}, we add an error term ($\xi$)  with a zero mean and a standard deviation of $\sigma_\xi$ to model a continuous (Gaussian) likelihood. The likelihood of an observed outcome $Y=Y_i$ at input $\mathbf{X}_i$ and $A_i$ is calculated using a Gaussian distribution with parameters $\zeta_i$ and $\sigma_\xi$. Here, $\sigma_\xi$ becomes an additional  parameter inferred from data in addition to the model parameters ($\boldsymbol{\theta_Y}$). Since $\sigma_\xi$ is a positive quantity (it being a standard deviation), we assume a distribution such as Half Normal that is defined only on positive values with a parameter $\sigma_o$, which can be replaced with a suitable value as needed. In Equations \ref{eqn:bintbino_prop}, \ref{eqn:bintbino_out} and \ref{eqn:bintconto_out}, the HalfNormal, Normal, and Multivariate Gaussian (MVN) priors are used for illustration purposes only, and can be replaced with appropriate priors as needed.

Approximate Bayesian inference techniques such as Markov Chain Monte Carlo (MCMC) \citep{chib2001markov}  and Stochastic Variational Inference (SVI) methods \citep{hoffman2013stochastic} will yield samples from the posterior distributions of model parameters, which are then used to predict outcomes and treatment probabilities. If $d$ and $m$ represent the number of data points, and the number of posterior samples respectively, then outcome model predictions and clever covariate terms ($H$), calculated using Equation \ref{eqn:clever}, will  be $d \times m$ matrices. Following the outcome model predictions and clever covariate terms, the next step is training the fluctuation model formulated in Equation \ref{eqn:fluctuation}.

\subsubsection{Fluctuation model}
\label{subsubsec:fluc_model}

Similar to the outcome model above, we describe the fluctuation model formulations for binary and continuous outcome separately. For a binary outcome, we use the model formulation given in Equation \ref{eqn:fluctuation}, and for a continuous outcome, we use a linear regression with a Gaussian likelihood as shown below in Equation \ref{eqn:cont_fluctuation2} \citep{van2006targeted, gruber2012tmle}.

\begin{equation}
\label{eqn:cont_fluctuation2}
    Y = Y_{init} + \epsilon H + \xi
\end{equation}

In Equation \ref{eqn:cont_fluctuation2}, $\xi$ is the error term modeled using a Gaussian distribution. A Gaussian likelihood is chosen here but it can be changed as desired. The estimation of fluctuation model parameters ($\epsilon$), in the presence of a binary or continuous outcome, is not straightforward as the true outcome ($Y$) is a $d\times 1$ vector, and as mentioned above, the initial outcome model predictions ($Y_{init}$) and clever covariate terms ($H(A, \mathbf{X})$) are $d\times m$ matrices. We propose two approaches below to alleviate this problem.

\underline{\textit{Approach 1:}} The key idea in this approach is to use \underline{only} the mean values of the initial predictions and the clever covariate values to obtain the posterior distributions of fluctuation parameters. By using the mean values, $Y_{init}$ and $H(A, \mathbf{X})$, which are initially $d\times m$ matrices now become d-dimensional vectors (or $d\times 1$ matrices). The formulation of the fluctuation parameter model using the mean values is shown in Equation \ref{eqn:mean_fluctuation}.

\begin{equation}
\label{eqn:mean_fluctuation}
    \mbox{logit}(Y) = \mbox{logit}(\mathop{{}\mathbb{E}_{\text{col}}}[Y_{init}]) + \epsilon \mathop{{}\mathbb{E}_{\text{col}}}[H(A, \mathbf{X})] 
\end{equation}
In Equation \ref{eqn:mean_fluctuation}, $\mathop{{}\mathbb{E}_{\text{col}}}[Y_{init}]$ and $\mathop{{}\mathbb{E}_{\text{col}}}[H(A, \mathbf{X})]$ are the mean values of initial predictions and clever covariate terms calculated as $(\mathop{{}\mathbb{E}_{\text{col}}}[Z])_k  = \frac{1}{m} \sum_{l=1}^m Z_{kl}, Z\in \{Y_{init}, H(A, \mathbf{X})\}$, where $k$ and $l$ represent matrix row and column indices. Equation \ref{eqn:bayes_fluc1} shows the Bayesian formulation of the above model.

\begin{equation}
\label{eqn:bayes_fluc1}
\begin{split}
\epsilon & \sim \mbox{Normal}(\mu_{\epsilon}, \sigma_{\epsilon})\\
\kappa_i &= \mbox{expit}(\mbox{logit}(\mathop{{}\mathbb{E}}[Y_{init,i}]) + \epsilon \mathop{{}\mathbb{E}}[H_i] )\\
Y_i|\epsilon, \mathop{{}\mathbb{E}}[Y_{init,i}], \mathop{{}\mathbb{E}}[H_i] & \sim \mbox{Bernoulli}(\kappa_i)\\
p(\epsilon|Y,\mathop{{}\mathbb{E}}[Y_{init}], \mathop{{}\mathbb{E}}[H]) & \propto p(\epsilon) \prod_{i=1}^d p(Y_i|\epsilon, \mathop{{}\mathbb{E}}[Y_{init,i}], \mathop{{}\mathbb{E}}[H_i])
\end{split}
\end{equation}

In the presence of a continuous outcome, Equations \ref{eqn:mean_fluctuation} and \ref{eqn:bayes_fluc1} are modified as shown below in Equations \ref{eqn:mean_cont_fluctuation} and \ref{eqn:bayes_cont_fluc1} respectively.

\begin{equation}
\label{eqn:mean_cont_fluctuation}
    Y = \mathop{{}\mathbb{E}_{\text{col}}}[Y_{init}] + \epsilon \mathop{{}\mathbb{E}_{\text{col}}}[H] + \xi
\end{equation}

\begin{equation}
\label{eqn:bayes_cont_fluc1}
\begin{split}
\sigma_\xi &\sim \mbox{HalfNormal}(\sigma_f) \\
\xi|\sigma_\xi &\sim \mbox{Normal}(0, \sigma_\xi)\\
\epsilon & \sim \mbox{Normal}(\mu_{\epsilon}, \sigma_{\epsilon})\\
\kappa_i &= \mathop{{}\mathbb{E}}[Y_{init}]_i +   \epsilon \mathop{{}\mathbb{E}}[H]_i\\
Y_i|H_i, \epsilon, & Y_{init,i}, \sigma_\xi  \sim \mbox{Normal}(\kappa_i, \sigma_\xi)\\
p(\epsilon, \sigma_\xi|Y,H,Y_{init}) & \propto p(\epsilon) p(\sigma_\xi) \prod_{i=1}^d p(Y_i|H_i, \epsilon, Y_{init,i}, \sigma_\xi)
\end{split}
\end{equation}

Similar to the Bayesian propensity and outcome model formulations above, we begin with the prior distributions over the model parameters followed by the likelihood equation; the prior and likelihood expressions are used to obtain posterior distributions. In this paper, we term this Bayesian implementation of TMLE where only mean values of initial outcome and clever covariate predictions are used for fluctuation model training as \underline{\textbf{B}}ayesian \underline{\textbf{TMLE}} with \underline{\textbf{M}}ean predictions (B-TMLE-M). A key drawback of this approach is we disregard the uncertainty in the initial model predictions and clever covariate terms leading to inaccurate posterior distribution of fluctuation parameter resulting in inaccurate model predictions, and thus inaccurate uncertainty quantification in estimated causal estimands.

\underline{\textit{Approach 2:}} In the second approach, we use not just the mean values but also the standard deviations of the initial outcome and clever covariate predictions. As discussed above, $Y_{init}$ and $H(A, \mathbf{X})$ are $d\times m$ matrices, i.e., at each of the $d$ data points, there are $m$ samples using which we compute the mean and standard deviation of predictions. Let $\mu_{Y_{init}}$ and $\sigma_{Y_{init}}$ represent the mean and standard deviation values of initial outcome predictions, and  $\mu_{H}$ and $\sigma_{H}$ represent mean and standard deviation values of clever covariate predictions. Since there are $d$ data points, each of these is a vector with $d$ values, i.e., a matrix dimension of $d \times 1$.

We model this scenario as Bayesian model training with input uncertainty where the true values of inputs (initial outcome and clever covariate predictions) are unknown but noisy observations are available \citep{huard2006bayesian}. In this formulation, we infer the true (latent) values of initial outcome and clever covariate predictions along with the fluctuation parameter ($\epsilon$). In total, we infer $d$ initial outcome predictions, $d$ clever covariate predictions, and the fluctuation parameter $\epsilon$ resulting in $2d+1$ parameters, where $d$ represents the number of data points. We provide below the corresponding Bayesian fluctuation model formulation with a binary outcome in Equation \ref{eqn:bayes_fluc_binary_s}.

\begin{equation}
\label{eqn:bayes_fluc_binary_s}
\begin{split}
\epsilon & \sim \mbox{Normal}(\mu_\epsilon, \sigma_\epsilon)\\
Y_{init,i}|\mu_{Y_{init,i}}, \sigma_{Y_{init,i}} &\sim \mbox{Normal}(\mu_{Y_{init,i}}, \sigma_{Y_{init,i}})\\
H_i|\mu_{H_i}, \sigma_{H_i} &\sim \mbox{Normal}(\mu_{H_i}, \sigma_{H_i})\\
\kappa_i &= Y_{init,i} + \epsilon H_i\\
Y_i|Y_{init,i}, H_i &\sim \mbox{Bernoulli}(\kappa_i)\\
p(Y_{init}, H, \epsilon|Y, \mu_{Y_{init}}, \sigma_{Y_{init}}, \mu_{H}, \sigma_{H})& \\
\propto p(\epsilon)\prod_{i=1}^d p(Y_i|Y_{init,i}, H_i, \epsilon) p(Y_{init,i}&|\mu_{Y_{init,i}}, \sigma_{Y_{init,i}})p(H_i|\mu_{H_i}, \sigma_{H_i})
\end{split}
\end{equation}

In Equation \ref{eqn:bayes_fluc_binary_s}, we assume that at each data point, the true initial outcome and clever covariate predictions are unknown, and therefore become inference parameters in the Bayesian paradigm. Since prior distributions are required in Bayesian paradigm for inference parameters, we assume Gaussian (or Normal) priors with available mean and standard deviation values as prior distribution parameters.  In the presence of a continuous outcome, we also infer the standard deviation ($\sigma_\xi$) with respect to the measurement error ($\xi$).  The corresponding Bayesian fluctuation model formulation with a continuous outcome is given below in Equation \ref{eqn:bayes_fluc_cont_s}.

\begin{equation}
\label{eqn:bayes_fluc_cont_s}
\begin{split}
\sigma_\xi &\sim \mbox{HalfNormal}(\sigma_o) \\
\xi|\sigma_\xi &\sim \mbox{Normal}(0, \sigma_\xi)\\
\epsilon & \sim \mbox{Normal}(\mu_\epsilon, \sigma_\epsilon)\\
Y_{init,i}|\mu_{Y_{init,i}}, \sigma_{Y_{init,i}} &\sim \mbox{Normal}(\mu_{Y_{init,i}}, \sigma_{Y_{init,i}})\\
H_i|\mu_{H_i}, \sigma_{H_i} &\sim \mbox{Normal}(\mu_{H_i}, \sigma_{H_i})\\
\kappa_i &= Y_{init,i} + \epsilon H_i\\
Y_i|Y_{init,i}, H_i &\sim \mbox{Normal}(\kappa_i, \sigma_\xi)\\
p(Y_{init}, H, \epsilon, \sigma_\xi|Y, \mu_{Y_{init}}, \sigma_{Y_{init}}, \mu_{H}, \sigma_{H})& \\
\propto p(\epsilon) p(\sigma_\xi)\prod_{i=1}^d p(Y_i|Y_{init,i}, H_i, \epsilon) p(Y_{init,i}&|\mu_{Y_{init,i}}, \sigma_{Y_{init,i}})p(H_i|\mu_{H_i}, \sigma_{H_i})
\end{split}
\end{equation}

In this paper, we term this Bayesian implementation of TMLE where both the mean and standard deviation values of initial outcome and clever covariate predictions are used for fluctuation model training as \underline{\textbf{B}}ayesian \underline{\textbf{TMLE}} with \underline{\textbf{S}}ummary \underline{\textbf{S}}tatistics (B-TMLE-SS).  Here, we considered the available mean and standard deviation values as summary statistics of the initial outcome and clever covariate predictions. In the above formulation, we used only the mean and standard deviation values as summary statistics; however, additional summary statistics such as skewness and kurtosis (third and fourth moments) with appropriate prior distributions can also be used.  We should note that in both B-TMLE-M and B-TMLE-SS, the training procedures for the outcome and propensity models are the same, and they differ only in the formulation and training of fluctuation model. Both these formulations implement sequential Bayesian training of the three models in the TMLE framework.  We illustrated the  B-TMLE-M and B-TMLE-SS algorithms using a one-parameter fluctuation model formulation; however, the one-parameter formulation can easily be replaced with a two-parameter fluctuation model formulation discussed in Equation \ref{eqn:fluctuation2}. 

\subsection{Bayesian Network implementation of TMLE (BN-TMLE)}
\label{subsec:bn-tmle}

\begin{figure}[!htb]
    \centering
    \includegraphics[width=0.42\linewidth]{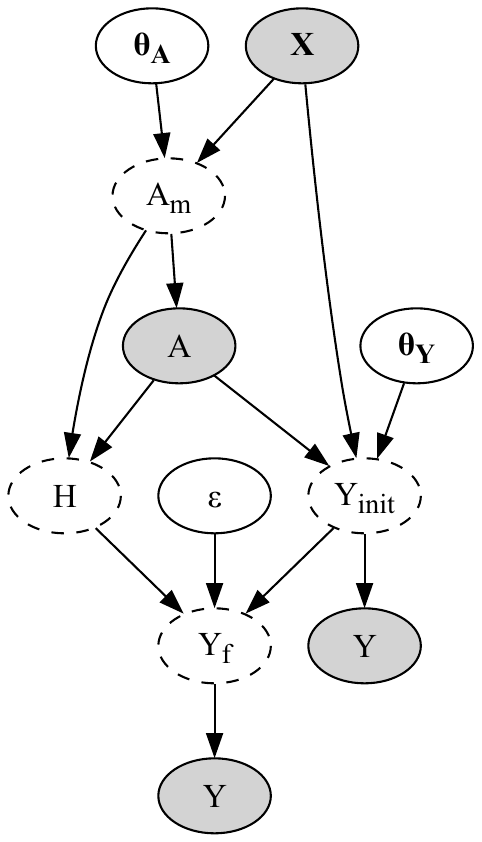}
    \caption{Bayesian network combining the outcome, treatment, and fluctuation models of the TMLE framework for causal effect estimation for a binary outcome}
    \label{fig:bn_tmle}
\end{figure}

In this third Bayesian implementation of the TMLE algorithm,  we use a Bayesian network to combine the three models (outcome, propensity, and fluctuation), and train them simultaneously.  A brief introduction to Bayesian networks and their learning are provided in \ref{app:bn} and \ref{app:learn_bn} respectively. Figure \ref{fig:bn_tmle} shows an illustrative Bayesian network for a binary outcome, where shaded nodes represent the nodes on which data are available (observation nodes). Here, the shaded nodes are the confounders, treatment and outcome variables ($\mathbf{X}, A, Y$).  Nodes with dashed borders are deterministic nodes while the remaining nodes represent inference parameters to be estimated from data.  The Bayesian network is drawn following the qualitative dependence relationships between several variables (such as confounders, treatment and outcome variables).

Given the generic formulation of the propensity model in Equation \ref{eqn:propensity}, the conditional distribution of the treatment conditioned on the input and model parameters can be written as shown in Equation \ref{eqn:prop_gen}. 

\begin{equation}
\label{eqn:prop_gen}
    \begin{split}
       A|\mathbf{X}, \boldsymbol{\theta_A} &\sim \mbox{Bernoulli}(\mbox{expit}(h(\mathbf{X}; \boldsymbol{\theta_A})))
    \end{split}
\end{equation}

The propensity score prediction at any input denoted by $A_m|\mathbf{X}, \boldsymbol{\theta_A}$ can be calculated as $A_m|\mathbf{X}, \boldsymbol{\theta_A} = \mbox{expit}(h(\mathbf{X}; \boldsymbol{\theta_A}))$. This is because the expected (mean) value of a Bernoulli random variable is its parameter value, which is the probability of success. Therefore, the conditional distribution in Equation \ref{eqn:prop_gen} can be written as $p(A|A_m) = \mbox{Bernoulli}(A_m)$. $A_m$ is a deterministic value given $\mathbf{X}$ and $\boldsymbol{\theta_A}$ and thus represented as a deterministic node in Figure \ref{fig:bn_tmle}. Given the treatment value and the propensity score, the clever covariate can be deterministically calculated using Equation \ref{eqn:clever}
 and thus represented as a deterministic node. 
 
 Similarly, the mean prediction of the outcome variable ($Y_{init}$) using the outcome model in Equation \ref{eqn:outcome}, and associated conditional distribution of the outcome variable can be written as shown below in Equation \ref{eqn:out_gen}.

\begin{equation}
\label{eqn:out_gen}
    \begin{split}
    Y_{init}|\mathbf{X}, A, \boldsymbol{\theta_Y} &= \mbox{expit}(g(\mathbf{X}, A; \boldsymbol{\theta_Y}))\\
    Y|Y_{init} &\sim \mbox{Bernoulli}(Y_{init})
    \end{split}
\end{equation}

Given the mean prediction (probability of success), the outcome variable $Y$ follows a Bernoulli distribution with a parameter value of $Y_{init}$.  The initial outcome predictions and clever covariate values are then used, along with the fluctuation parameter $\epsilon$ , to train the fluctuation model shown in Equation \ref{eqn:fluctuation}.  The mean prediction from the fluctuation model, which is the updated outcome prediction, using the initial outcome prediction and the clever covariate values, and associated conditional distribution of the outcome variable are given below in Equation \ref{eqn:fluc_gen}.

\begin{equation}
\label{eqn:fluc_gen}
    \begin{split}
        Y_f|Y_{init}, H, \epsilon &= \mbox{expit}(\mbox{logit}(Y_{init}) + \epsilon H)\\
        Y|Y_f &\sim \mbox{Bernoulli}(Y_f)
    \end{split}
\end{equation}

Note that since the same outcome data ($Y$) is used for training the outcome and fluctuation models, we duplicate the node $Y$ so we can train the outcome and fluctuation model directly.  After formulating all the conditional distributions, the available data on confounders, treatment and outcome variables are used to infer the parameters of all three models ($\boldsymbol{\theta_Y}, \boldsymbol{\theta_A}, \epsilon$) simultaneously. We should note here that unlike the the first two approaches which used the mean values and summary statistics of initial outcome and clever covariates, the integration of all models through a Bayesian network enables to train the fluctuation model using all the samples from $Y_{init}$ and $H$ providing more accurate updated predictions of the outcome variable when compared to the first two approaches.

\begin{figure}[htbp]
    \centering
    \includegraphics[width=0.5\linewidth]{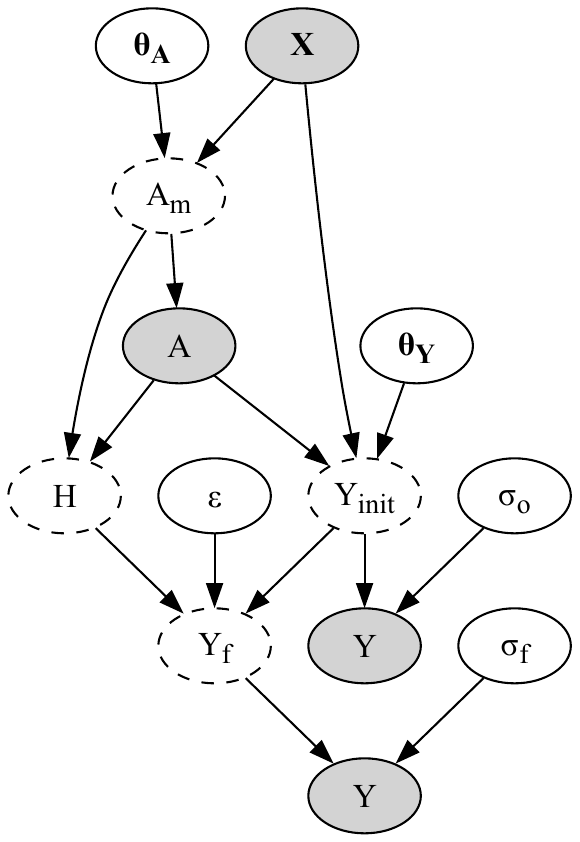}
    \caption{Bayesian network combining the outcome, treatment, and fluctuation models of the TMLE framework for causal effect estimation for a continuous outcome}
    \label{fig:bn_tmle_cont}
\end{figure}

Next, we extend the above Bayesian network discussion for a continuous outcome. Figure \ref{fig:bn_tmle_cont} shows the Bayesian network for a continuous outcome. A key difference in this Bayesian network is the presence of additional inference parameters, specifically, the standard deviations of the Gaussian error terms associated with the continuous outcome when training the outcome and fluctuation models denoted as $\sigma_o$ and $\sigma_f$ for outcome and fluctuation models respectively. The expressions for $A_m, A$ and $H$ will remain the same as discussed above for the case of a binary outcome. The initial outcome prediction $Y_{init}$ and corresponding outcome conditional distribution are given below in Equation \ref{eqn:out_gen_cont}.

\begin{equation}
\label{eqn:out_gen_cont}
    \begin{split}
    Y_{init}|\mathbf{X}, A, \boldsymbol{\theta_Y} &= g(\mathbf{X}, A; \boldsymbol{\theta_Y})\\
    Y|Y_{init}, \sigma_o &\sim \mbox{Normal}(Y_{init}, \sigma_o)
    \end{split}
\end{equation}

Similarly, the updated outcome prediction $Y_f$ and associated outcome conditional distribution are given below in Equation \ref{eqn:fluc_gen_cont}.

\begin{equation}
\label{eqn:fluc_gen_cont}
    \begin{split}
       Y_f|Y_{init}, H, \epsilon &= Y_{init} + \epsilon H\\
       Y|Y_f, \sigma_f &\sim \mbox{Normal}(Y_f, \sigma_f)
    \end{split}
\end{equation}

\begin{figure}[!htb]
    \centering
    \subfigure[Binary outcome]{
        \includegraphics[width=0.43\textwidth]{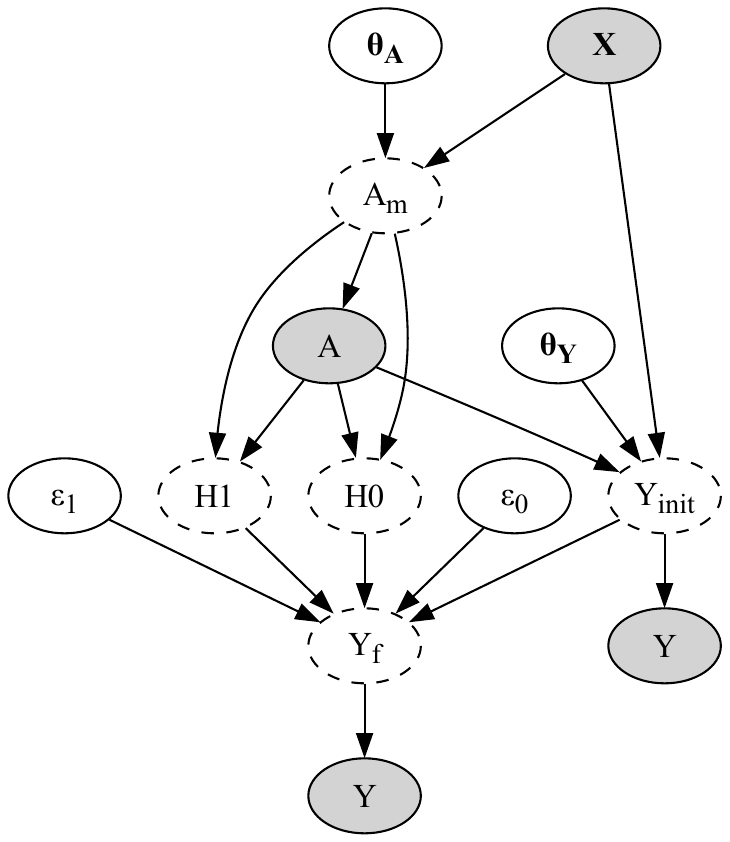} 
    }
    \subfigure[Continuous outcome]{
        \includegraphics[width=0.52\textwidth]{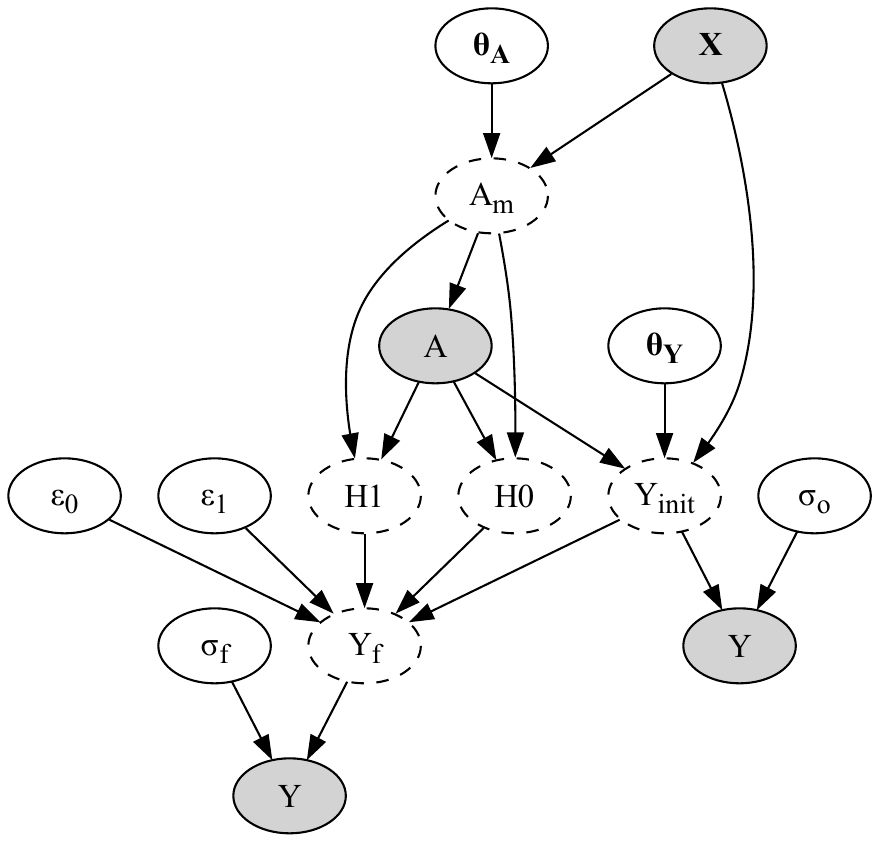}     
    }
    \vspace{0.5cm}

    \caption{Bayesian network TMLE formulations considering two-parameter fluctuation model formulations for binary and continuous outcomes}
    \label{fig:bn2-tmle}
\end{figure}

The BN-TMLE formulation discussed above considered a one-parameter fluctuation model formulation, which can easily be replaced with a two-parameter fluctuation model formulation. Figure \ref{fig:bn2-tmle} shows Bayesian networks with two-parameter fluctuation formulations for both binary and continuous outcomes. As discussed in Section \ref{subsec:tmle}, the two-parameter formulation results in an additional inference parameter when compared to the one-parameter formulation.

\subsection{Uncertainty quantification in causal estimands}
\label{subsec:uq}

After estimating the parameters associated with all the three models using any of the three approaches (B-TMLE-M, B-TMLE-SS and BN-TMLE), the next step is the estimation of the Average Treatment Effect (ATE) and associated uncertainty. Following the Potential Outcomes Framework (POF) (POF), we need to calculate the outcome predictions with and without the treatment ($A=0$ and $A=1$).

In the two SB-TMLE approaches (B-TMLE-M and B-TMLE-SS), we follow a sequential approach for estimating updated model predictions. For the available $d$ data points, we calculate initial outcome predictions with and without treatment($Y_{init}(1, \mathbf{X})$ and $Y_{init}(0, \mathbf{X})$) and treatment probabilities ($P(A=1|\mathbf{X})$ and $P(A=0|\mathbf{X})$) as $d \times m$ matrices, i.e., $m$ realizations at each input corresponding to the $m$ posterior samples of model parameters in the outcome and propensity models. Using probabilistic estimates of treatment probabilities, we compute probabilistic estimates of the clever covariate, $H(A, \mathbf{X})$, as a $d \times m$ matrix. Using the probabilistic estimates of initial predictions, clever covariates, and posterior samples of fluctuation parameters, we obtain updated probabilistic estimates of outcome with and without treatment ($Y_f(1, \mathbf{X})$ and $Y_f(0, \mathbf{X})$) as $d\times m$ matrices using either the B-TMLE-M or the B-TMLE-SS approach. 

Using the BN-TMLE approach, the updated model predictions ($Y_f$) with and without treatment can be obtained as $d\times m$ matrices by intervening the system and setting $A=1$ and $A=0$ respectively. Mathematically, the interventions are implemented using do operations \citep{pearl1994probabilistic} by setting $\mbox{do}(A=1)$ and $\mbox{do}(A=0)$ respectively. We obtain interventional distributions of $Y_f$, i.e., distributions of $Y_f$ after setting $\mbox{do}(A=a), a \in {0,1}$ in three steps: \textbf{(1)} Remove all incoming edges to the treatment variable ($A$), \textbf{(2)} set the treatment variable to a desired value $A=0,1$, and \textbf{(3)} obtain the distribution of $Y_f$ using data on confounders ($X$), treatment variable set at the desired value, and posterior distributions of the inferred model parameters associated with the outcome, propensity and fluctuation models.

Using the two $d \times m$ matrices of updated model predictions with and without treatment, we calculate their expected values with and without treatment across data points through a column-wise mean computation resulting in two arrays with $m$ elements each with and without treatment. Using these $m$ values with and without treatment, we can compute $m$ values of the ATE, which can later used to construct a probability distribution and credible intervals of the ATE. This probability distribution represents the uncertainty associated with ATE due to the uncertainty in the model parameters. A credible interval represents a range (derived from the probability distribution) that contains the variable of interest (ATE) with a specified probability. For example, a 95\% credible interval of ATE is defined as a range that has a 95\% probability for the ATE to lie within.

\section{Experiments}
\label{sec:exp}

In this section, we present three sets of experiments to demonstrate the proposed Bayesian TMLE approaches and compare them against classical TMLE implementations. Section \ref{subsec: case} compares the performance of the three Bayesian TMLE approaches (B-TMLE-M, B-TMLE-SS and BN-TMLE) and classical implementation against the true ATE values for two simulated datasets with binary and continuous outcomes. Section \ref{subsec:data} investigates the effect of data size and also model misspecifications on Bayesian (BN-TMLE) and classical TMLE implementations, and compare their performance against the true ATE values on simulated datasets. Section \ref{subsec:fluc_exp} compares Bayesian TMLE implementations (specifically BN-TMLE) with one and two-parameter fluctuation formulations on the three datasets used in Section \ref{subsec: case}. Typically, we compute confidence intervals in classical TMLE implementations and credible intervals in proposed Bayesian approaches. Although they possess distinct definitions and interpretations, they are often similar in practice \citep{gray2015comparison}. For simplicity, we use the term confidence interval in both classical and Bayesian TMLE implementation in this section.

We used Numpyro probabilistic programming library in Python \citep{phan2019composable}, which is an extension to the popular Pyro library \citep{bingham2019pyro} for TMLE implementation in a Bayesian paradigm.  We used an MCMC-based inference, specifically the No-U-Turn  Sampler \citep{hoffman2014no} for Bayesian TMLE implementation. We used standard Gaussian distributions as priors for all parameters across all three models, and a HalfNormal distributions with a unit standard deviation as priors for the standard deviations of the Gaussian error terms associated with a continuous outcome. We ran two MCMC chains, each with a burn-in of 1000 samples and 2000 posterior samples, totaling to 4000 posterior samples across both the chains for each inference parameter.

\subsection{Case studies to demonstrate proposed approaches}
\label{subsec: case}

\begin{figure}[!htb]
    \centering
    \includegraphics[width=0.35\linewidth]{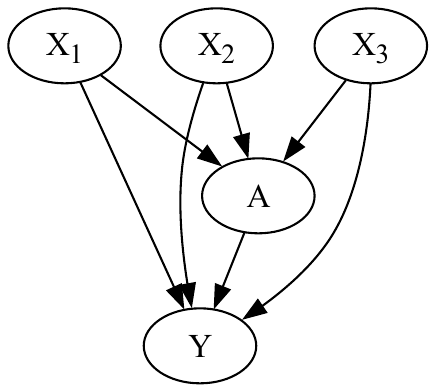}
    \caption{A directed acyclic graphical model (DAG) showing dependence between confounders, treatment and outcome variables for a simulated study to study the effect of data size and model misspecification}
    \label{fig:bn-simulation}
\end{figure}

As mentioned above, we generated two simulated case studies with binary and continuous outcomes to compare the performance of the three proposed Bayesian TMLE implementations against classical TMLE implementation and underlying true ATE values. For each case study, we considered a binary treatment ($A$) with three confounders ($X_1, X_2$ and $X_3$).  Figure \ref{fig:bn-simulation} provides a directed acyclic graph (DAG) model showing dependence relationships between  confounders, treatment and outcome variables. 

$X_1$ is a binary variable with two values 0 and 1, and modeled as a Bernoulli random variable with a success probability $(P(X_1=1))$ of 0.4. $X_2$ is a categorical random variable with three levels: 0, 1, and 2 with associated probabilities of 0.3, 0.5, and 0.2 respectively. $X_3$ is a continuous random variable modeled as a standard Gaussian distribution (mean and standard deviation of 0 and 1 respectively). Equation \ref{eqn:ps_A} is used for data generation of treatment variable. 

\begin{equation}
\label{eqn:ps_A}
    \begin{split}
        \mbox{logit}(A) &= -1.4 + 0.3X_1 + 0.5X_2 - 0.9X_3\\
        A &\sim \mbox{Bernoulli}(\mbox{expit}(\mbox{logit}(A)))
    \end{split}
\end{equation}

We used the same data generation process for the confounders and treatment variable but different expressions for binary and continuous outcome variable discussed below.

\underline{\textit{Binary outcome}}: Equation \ref{eqn:y0_Y} was used to generate data of a binary outcome using generated values of confounders and the treatment variable. Here, $\psi$ corresponds to the ATE, and we used a value of 0.03. In addition, random noise of 5\% was added to the outcome data. For each data point, we generated a uniform random value between 0 and 1. If the generated value is less than 0.05, we flipped the outcome value from 0 to 1 and vice versa.

\begin{equation}
\label{eqn:y0_Y}
    \begin{split}
        \mbox{logit}(Y_0) &= -1.3 - 0.7X_1 + 0.8X_2 + 0.9X_3\\
        &+ 0.07X_3^2 - 0.02X_1X_2 + 0.06X_2X_3\\
        P(Y_0 = 1) &= \mbox{expit}(\mbox{logit}(Y_0))\\
        P(Y_1 = 1) &= P(Y_0 = 1) + \psi\\
        Y & \sim \mbox{Bernoulli}(AP(Y_1 = 1) + (1-A)P(Y_0 = 1))
    \end{split}
\end{equation}

\underline{\textit{Continuous outcome}}: Equation \ref{eqn:y0_Y_cont} was used to generate data of a continuous outcome using generated values of confounders and the treatment variable. Here, $\psi$ corresponds to the ATE, and we used a value of 0.25.

\begin{equation}
\label{eqn:y0_Y_cont}
    \begin{split}
        Y_0 &= -1.3 - 0.7X_1 + 0.8X_2 + 0.9X_3\\
        &+ 0.07X_3^2 - 0.02X_1X_2 + 0.06X_2X_3\\
        Y_1 &= Y_0 + \psi\\
        Y & \sim \mbox{Normal}(AY_1 + (1-A)Y_0, 0.1)
    \end{split}
\end{equation}

For each of the two cases (binary and continuous outcomes), we generated 10000 data points and used them for causal effect estimation. Even though the same data generation processes were used to generate data on confounders and treatment variable, we used different random seeds for simulations resulting in different data values of confounders and treatment variable. 

\textbf{Data preprocessing and Model training}: As data preprocessing, we implemented one-hot encoding of the categorical variable ($X_2$), and normalization of continuous variables ($X_3$, and in the case of a continuous outcome, $Y$). For both outcome and treatment modeling, we identified Generalized linear models (such as logistic and linear regression with first order terms only) to produce the best fit after using Akaike Information Criterion (AIC) for model selection \citep{aho2014model}. One potential reason for this could be that AIC balances model fit (likelihood) and model complexity (number of model parameters), and a first-order model was favored since coefficients of the second order terms were much smaller when compared to the coefficients associated with the first order terms and a first order model has fewer model parameters. Also, we considered one-parameter fluctuation model formulation in both the case studies. We implemented the same data preprocessing steps and trained the same models for both classical and Bayesian TMLE analysis.  

\begin{figure}[!htbp]
    \centering
    \subfigure[Binary outcome]{
        \includegraphics[width=0.47\textwidth]
        {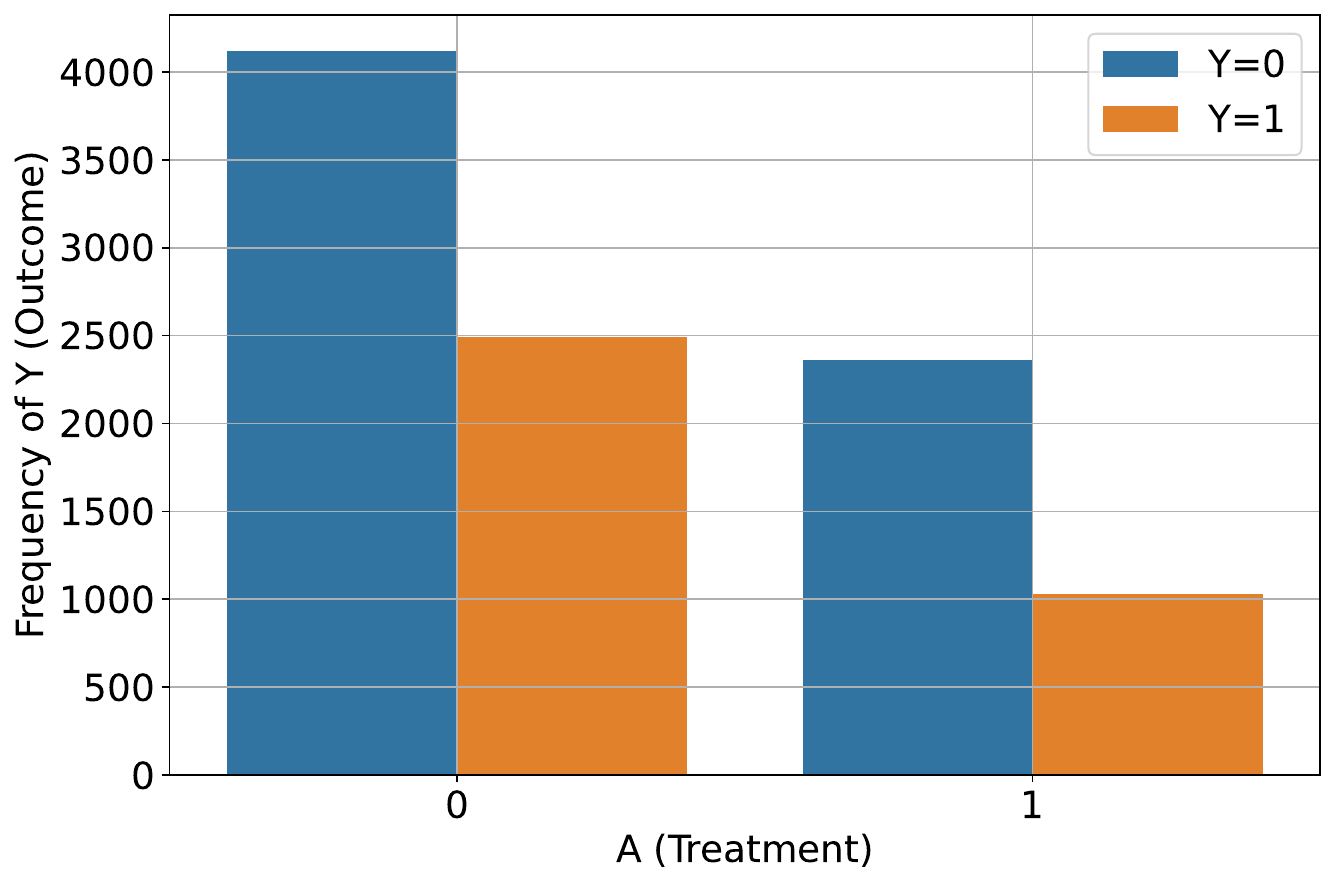}
        \label{fig:outcome_bin}
        
    }
    \subfigure[Continuous outcome]{
        \includegraphics[width=0.47\textwidth]
        {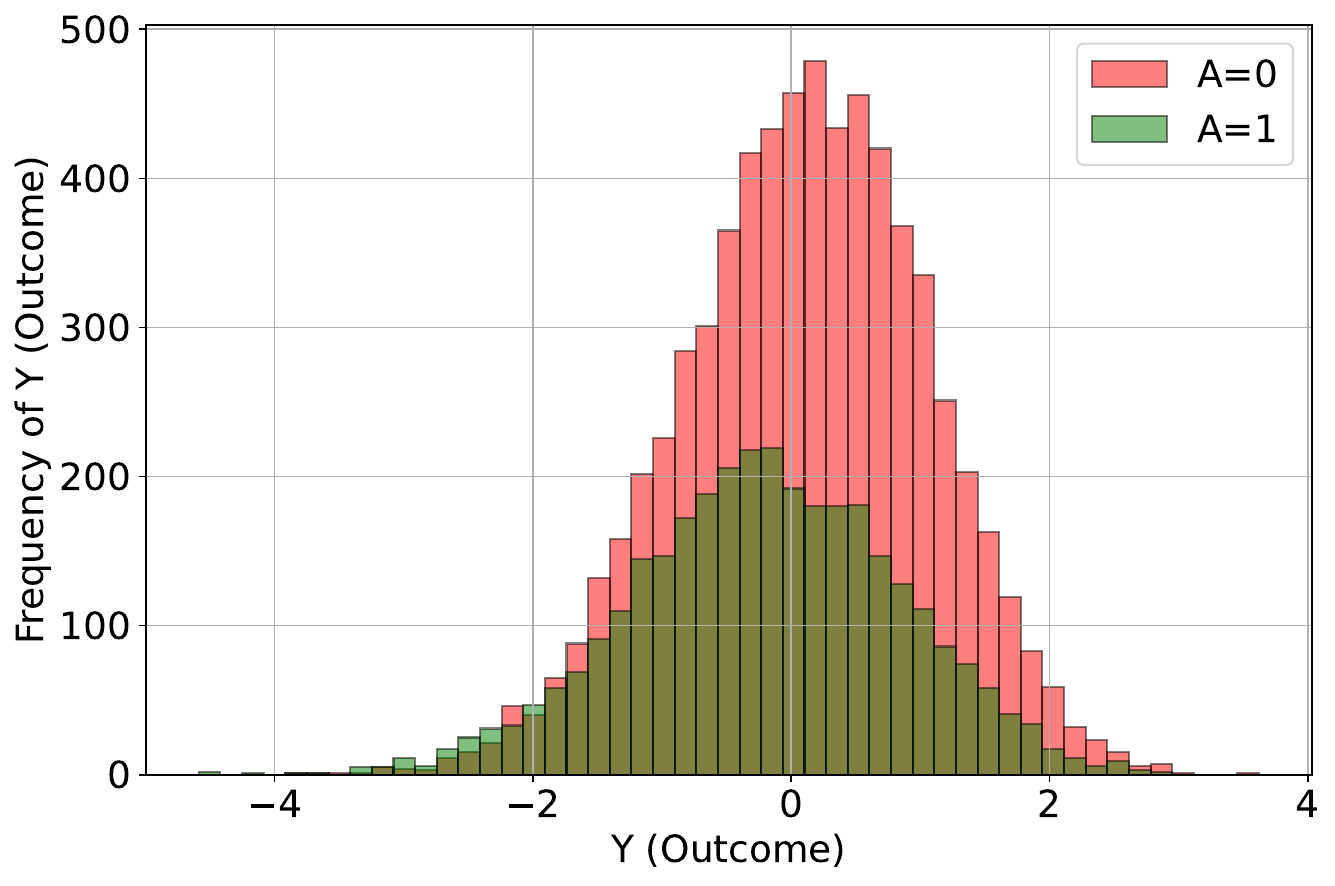}
        \label{fig:outcome_cont}
        
    }
    \caption{Distributions of outcome data across control and treatment groups for binary and continuous outcomes}
    \label{fig:output_dist}
\end{figure}

Figure \ref{fig:output_dist} shows the distributions of outcome data across the control and treatment groups ($A=0, 1$) for binary and continuous outcomes. From Figure \ref{fig:outcome_bin}, we can observe that there exists data points for both outcomes ($Y = 0, 1$) across both the control and treatment groups. Since the data generation process for the outcome variable ($Y$) follows a second order model, there exists model misspecification since we considered a first order model. Practical problems are often associated with model misspecification since it is not possible to identify the exact data generation process outcome and/or treatment variables. Here, we illustrated model misspecification in the outcome variable. Since TMLE is a Doubly Robust Estimation (DRE) estimation, it should be able to provide an accurate estimate of the underlying true ATE value when model misspecification exists in either the treatment or the outcome variable. 

\begin{figure}[!htbp]
    \centering
    \subfigure[Binary outcome]{
        \includegraphics[width=0.47\textwidth]
        {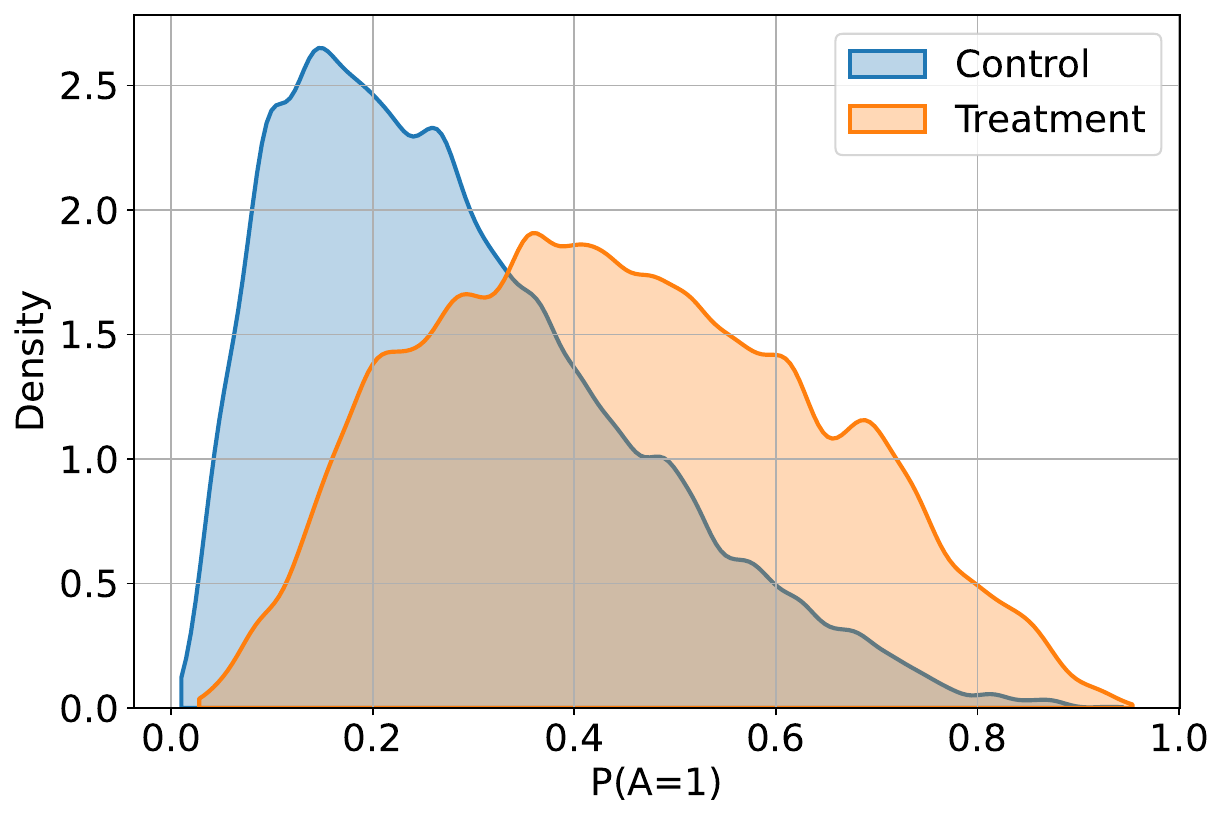}
        \label{fig:propensity_bin}
        
    }
    \subfigure[Continuous outcome]{
        \includegraphics[width=0.47\textwidth]
        {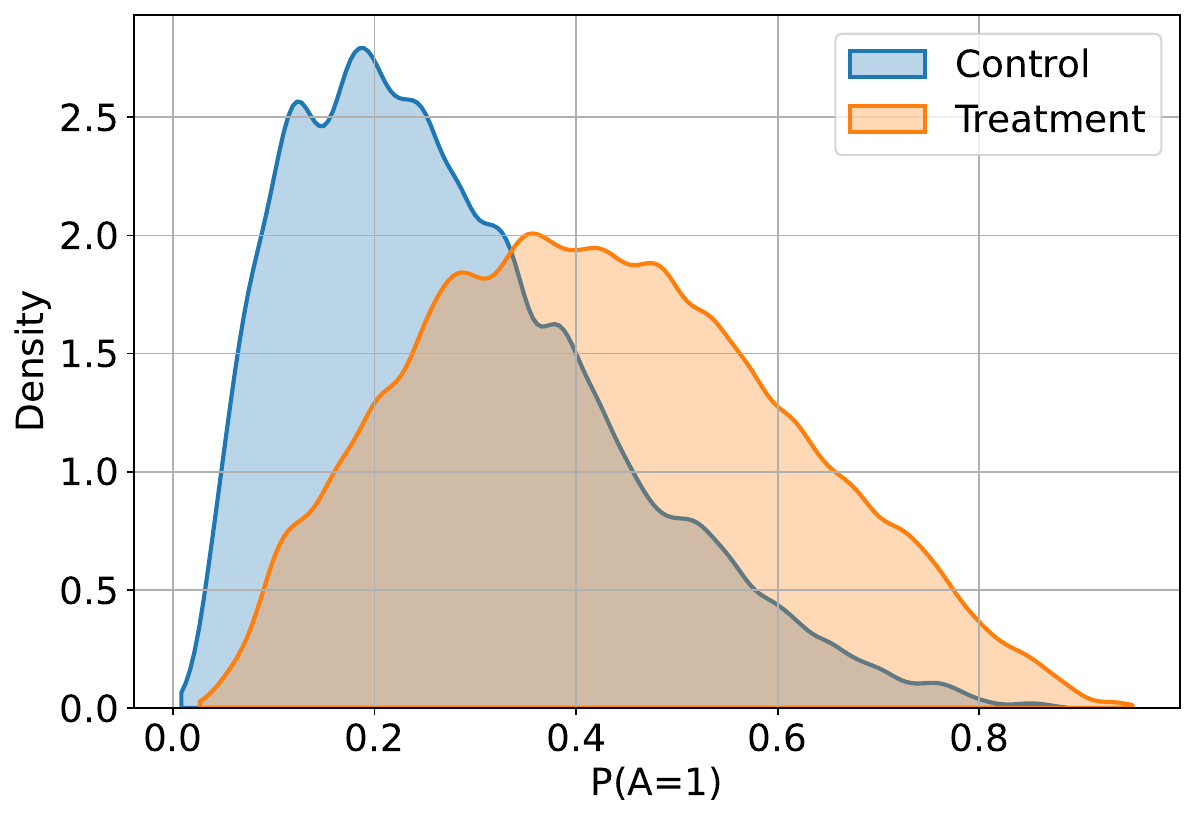}
        \label{fig:propensity_cont}
        
    }
    \caption{Distributions of propensity scores across control and treatment groups for binary and continuous outcomes}
    \label{fig:propensity_dist}
\end{figure}

Even though the data generation process for the treatment variable is the same for both binary and continuous outcomes (Equation \ref{eqn:ps_A}), as mentioned above, the data values are different since different random seeds were used for data generation. As mentioned above, we generated 4000 samples from the posterior distributions of model parameters resulting in 4000 samples of propensity scores at each of the 10000 data points. For visualization, we randomly selected 100 out of 4000 realizations of propensity scores and plotted kernel density estimation plots of the average of the 100 propensity score distributions, and plotted them in Figure \ref{fig:propensity_dist}, which shows that the distributions overlap at all propensity values across the control and treatment groups for both binary and continuous outcomes.

\begin{figure}[!htbp]
    \centering
    \subfigure[Binary outcome]{
        \includegraphics[width=0.47\textwidth]
        {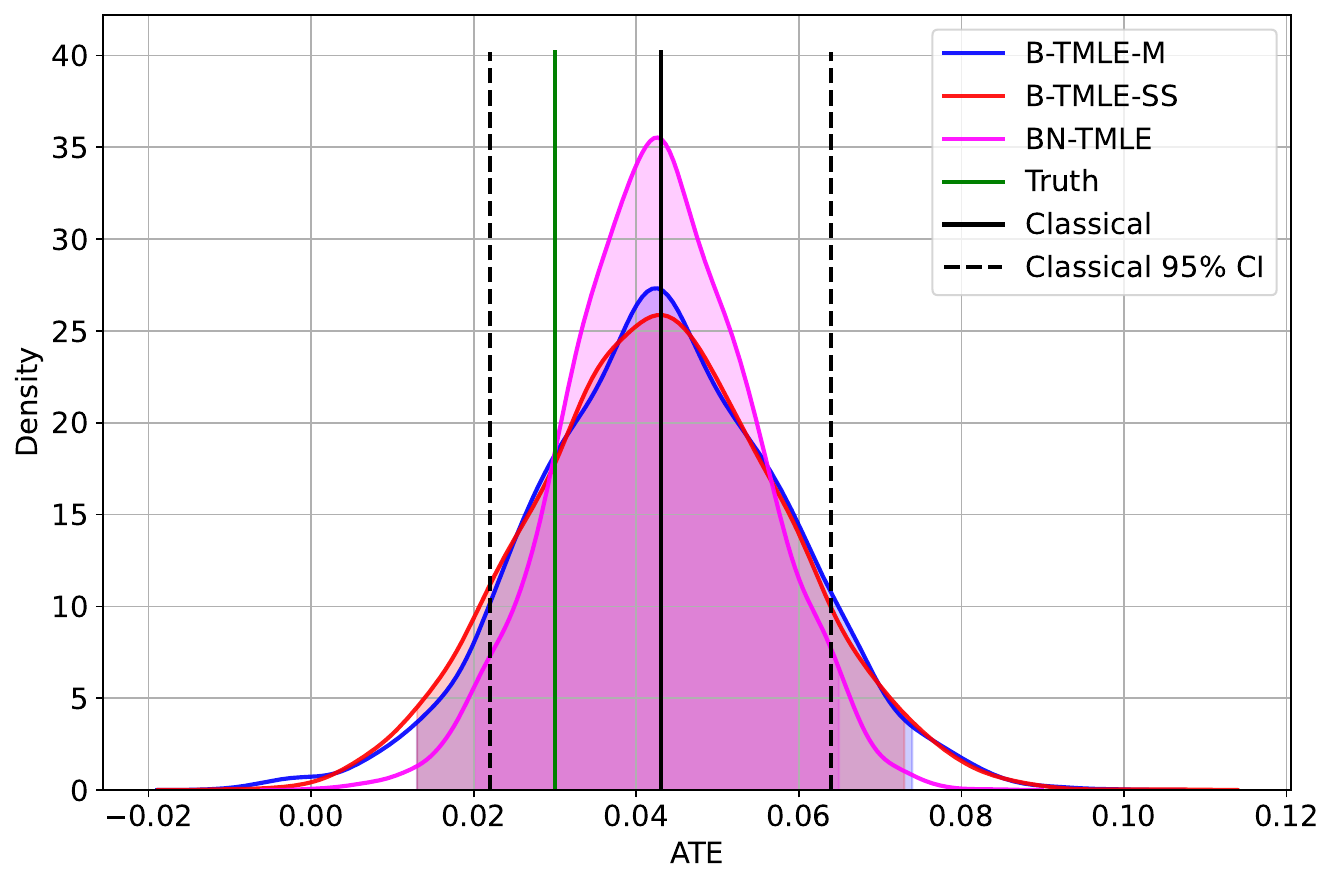}
        \label{fig:ate_bin}
        
    }
    \subfigure[Continuous outcome]{
        \includegraphics[width=0.47\textwidth]
        {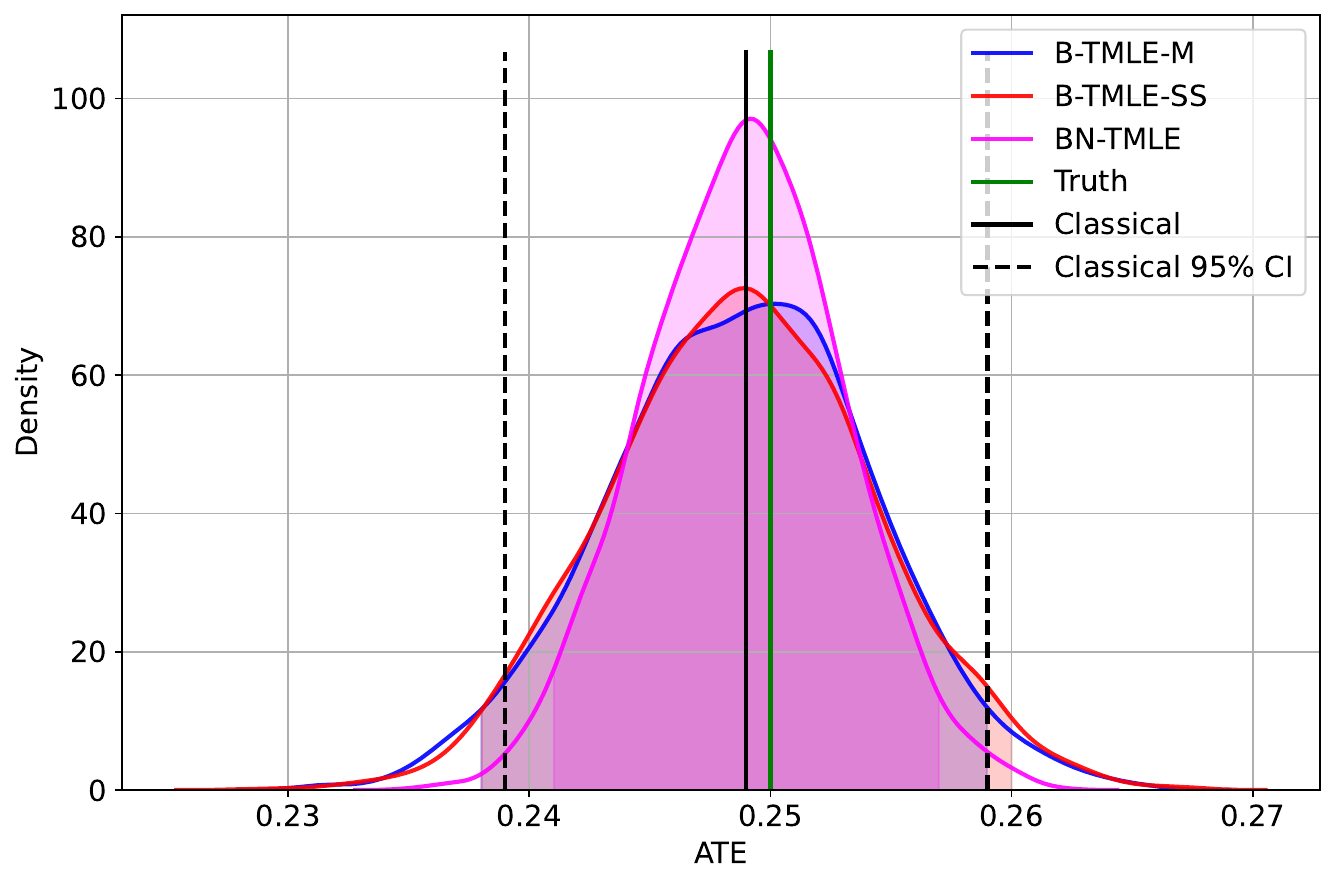}
        \label{fig:ate_cont}
        
    }
    \caption{Comparison of ATE distribution using the three Bayesian TMLE approaches (B-TMLE-M, B-TMLE-SS and BN-TMLE) against classical TMLE results and True values}
    \label{fig:ate_dist}
\end{figure}

\textbf{Causal Effect Estimation}: We obtained outcome predictions with and without treatments using all the three Bayesian approaches as two 10000 $\times$ 4000 matrices using which we computed 4000 samples of ATE. These 4000 samples were used to construct probability distribution of ATE using kernel density estimation; this distribution represents the uncertainty in the ATE estimate. Figure \ref{fig:ate_dist} compares the ATE distributions (along with 95\% confidence intervals) obtained using three Bayesian TMLE approaches and compared against classical TMLE implementation. The shaded regions below each of three distributions represent 95\% confidence intervals. Table \ref{tab:results} provides the mean and 95\% CI values of ATE from the three Bayesian approaches and classical implementation.

\begin{table}[htbp]
  \centering
  \caption{Comparing ATE Mean and 95\% CIs between proposed and classical methods}
  \vspace{0.3cm}
  \scalebox{0.68}{
    \begin{tabular}{l|c|c|c|c|c}
    \toprule
    \textbf{Outcome} & \textbf{Truth} & \textbf{Classical} & \textbf{B-TMLE-M} & \textbf{B-TMLE-SS} & \textbf{BN-TMLE} \\
    \midrule
    \textbf{Binary} & 0.03 & 0.043 [0.022, 0.064] & 0.043 [0.013, 0.074] & 0.043 [0.013, 0.073] & 0.043 [0.02, 0.065] \\
    \textbf{Continuous} & 0.25 & 0.249 [0.239, 0.259] & 0.249 [0.238, 0.259] & 0.249 [0.238, 0.26] & 0.249 [0.241, 0.257] \\
    \bottomrule
    \end{tabular}}
  \label{tab:results}
\end{table}

We can make the following observations from ATE comparison results in Figure \ref{fig:ate_dist} and Table \ref{tab:results}:
\begin{enumerate}
    \item The true ATE values are captured by the 95\% CIs of all three Bayesian approaches and the classical results.
    \item The mean ATE estimates from all the three Bayesian approaches and classical implementation are similar (even though they are slightly different from the true values). We believe the underlying difference is due to incorporation of random noise in the data generation process.
    \item  The ATE distributions associated with B-TMLE-M and B-TMLE-SS are similar, both in terms of mean and variance. The ATE distribution associated with BN-TMLE has a smaller variance (and narrower 95\% CIs) when compared to B-TMLE-M and B-TMLE-SS appraoaches. Of the three proposed Bayesian approaches, BN-TMLE provides the most accurate results.
    \item The 95\% CIs associated with BN-TMLE match well with those obtained from classical implementations with the Bayesian approach producing a slightly narrower CI in the case of the continuous outcome. The Bayesian approach also provides a full probability distribution, derived from posterior samples, as opposed to confidence intervals from classical implementation. 
\end{enumerate}

In Section \ref{subsec:data}, we studied the effect of data size and model misspecification on the ATE estimation performance of BN-TMLE and classical TMLE implementation.

\subsection{Effect of data size and model misspecification}
\label{subsec:data}

We considered 12 data sizes: 25, 50, 75, 100, 150, 200, 250, 300, 350, 400, 450, and 500. At each data size, we considered 100 replications, i.e., 100 datasets are randomly generated at each data size totaling to 1200 analyses (12 data sizes $\times$ 100 replications).

In addition to data size, we also studied the effect of model misspecification on causal effect estimation. Here, we define model misspecification as the discrepancy between the true data generating process and model. We studied three cases of model misspecification: (1) No model misspecification of outcome and propensity models, (2) model misspecification of the outcome model only, and (3) misspecification of both the outcome and propensity models. For each of the three cases above, we implemented the above data size analyses. In total, we have 3600 analyses (3 types of model misspecification cases $\times$ 12 data sizes $\times$ 100 replications). Finally, we repeated all the above analyses for two effect sizes resulting in a total of 7200 analyses (2 effect sizes $\times$ 3 types of model misspecification cases $\times$ 12 data sizes $\times$ 100 replications).

We re-used the data generation process in Section \ref{subsec: case} for generating data on confounders, treatment and outcome variables. For illustration, we considered binary outcome only to study the effect of data size and model misspecifications. In Section \ref{subsec: case}, we discussed that there exists model misspecification in the outcome variable. To study the effect of data size in the absence of model misspecifications, we generated treatment data using Equation \ref{eqn:ps_A} and the outcome data using a first-order model given in Equation \ref{eqn:y0_Y_first}. To study the effect of data size in the presence of model misspecification in outcome model, we generated treatment and outcome data using  Equations \ref{eqn:ps_A} and \ref{eqn:y0_Y} respectively.

\begin{equation}
\label{eqn:y0_Y_first}
    \begin{split}
        \mbox{logit}(Y_0) &= -1.3 - 0.7X_1 + 0.8X_2 + 0.9X_3\\
        P(Y_0 = 1) &= \mbox{expit}(\mbox{logit}(Y_0))\\
        P(Y_1 = 1) &= P(Y_0 = 1) + \psi\\
        Y & \sim \mbox{Bernoulli}(AP(Y_1 = 1) + (1-A)P(Y_0 = 1))
    \end{split}
\end{equation}

To study the effect of data size in the presence of model misspecifications in both treatment and outcome models, we used Equation \ref{eqn:ps_A_second} below for treatment data and Equation \ref{eqn:y0_Y_first} for outcome data.

\begin{equation}
\label{eqn:ps_A_second}
    \begin{split}
        \mbox{logit}(A) &= -1.4 + 0.3X_1 + 0.5X_2 - 0.9X_3\\
        &+ 0.07X_3^2 - 0.02X_1X_2 + 0.06X_2X_3\\
        A &\sim \mbox{Bernoulli}(\mbox{expit}(\mbox{logit}(A)))
    \end{split}
\end{equation}

 As discussed above, we considered two effect sizes ($\psi$) of 0.03 and 0.15. In all the three model misspecification cases, we considered logistic regression formulations for propensity and outcome models.  When model misspecification is considered and data generation followed a second-order process (Equations \ref{eqn:ps_A_second} and \ref{eqn:y0_Y}), we used a linear model (logistic regression) with first order terms as it had a lower Akaike Information Criterion (AIC) model selection score.  At each level of data size and model misspecification, 100 replications provide 100 mean values and 100 95\% CIs of ATE (using both Bayesian and classical TMLE implementations). Using these 100 values, we computed the mean and 95\% CI of the mean ATE value, and mean and 95\% CI of the width of the 95\% CI (difference between the upper and lower bounds of the 95\% CI). In addition, we also computed the coverage defined below in Equation \ref{eqn:coverage}. 

\begin{equation}
\label{eqn:coverage}
\mbox{Converage} = \dfrac{\mbox{\# replications truth is within predicted 95\% CI}}{\mbox{\# replications}} \times 100\%
\end{equation}

In addition to the coverage percentage, we also computed the corresponding 95\% CI. There are several approaches available to compute confidence intervals of proportions (or percentages) such as the Agresti-Coull interval, Clopper-Pearson interval, Wilson score interval, and Jeffreys Bayesian interval \citep{franco2019comparative}. Here, we used Jeffreys Bayesian interval for illustration.  

\begin{figure}[htbp]
    \centering
    \subfigure[NMS: Mean variation]{
        \includegraphics[width=0.3\textwidth]
        {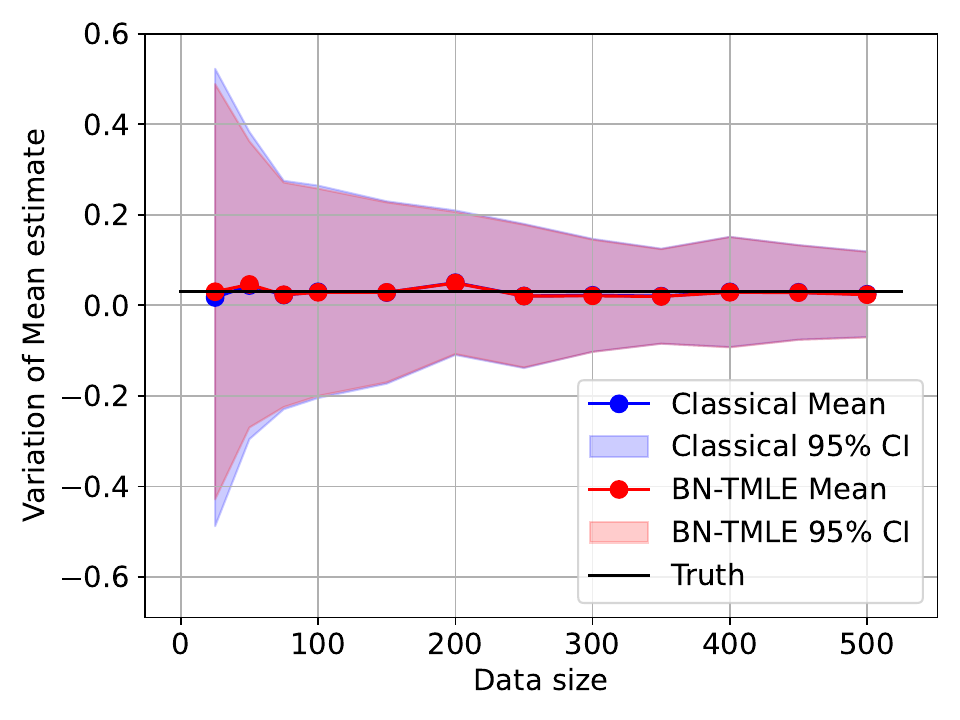}
        
    }
    \subfigure[NMS: Width variation]{
        \includegraphics[width=0.3\textwidth]
        {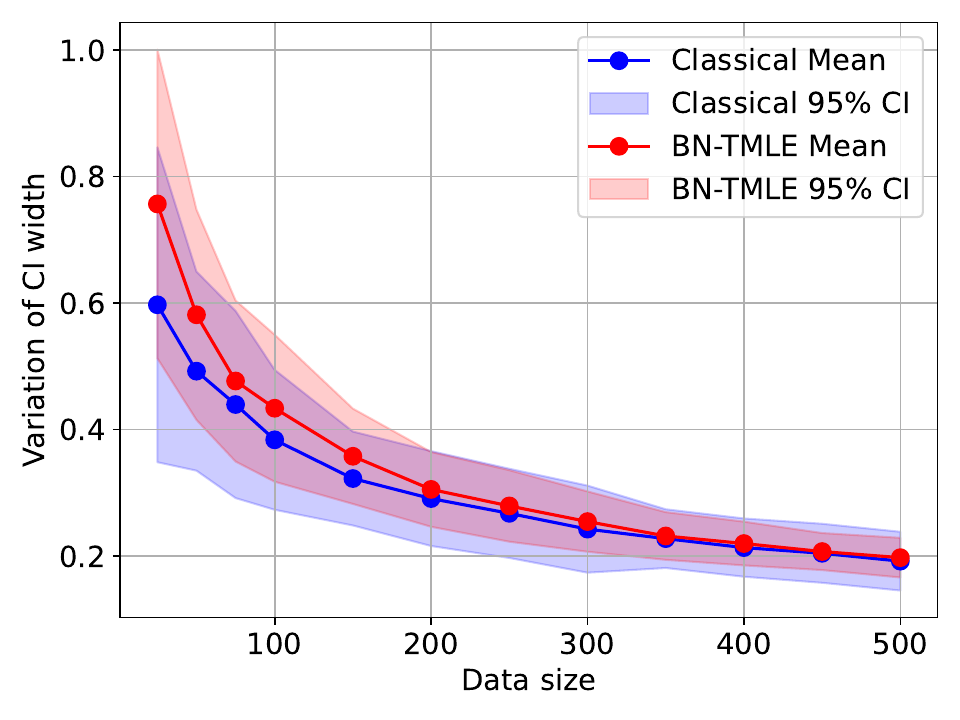}
        
    }
    \subfigure[NMS: Coverage]{
        \includegraphics[width=0.3\textwidth]
        {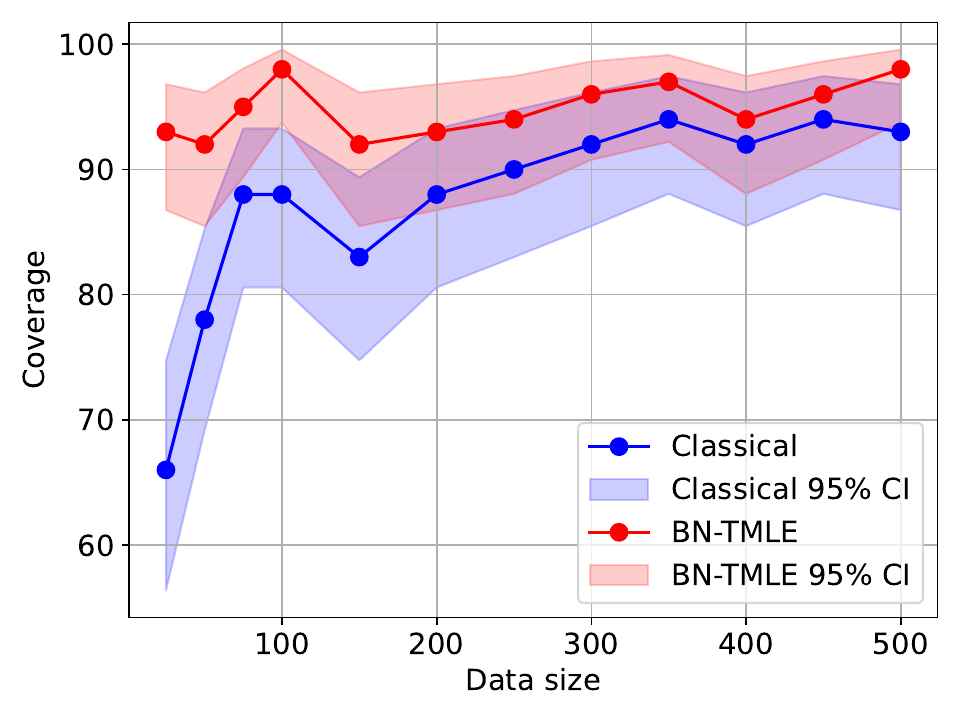}
        
    }
    \vspace{0.5cm}
    \subfigure[OMS: Mean variation]{
        \includegraphics[width=0.3\textwidth]
        {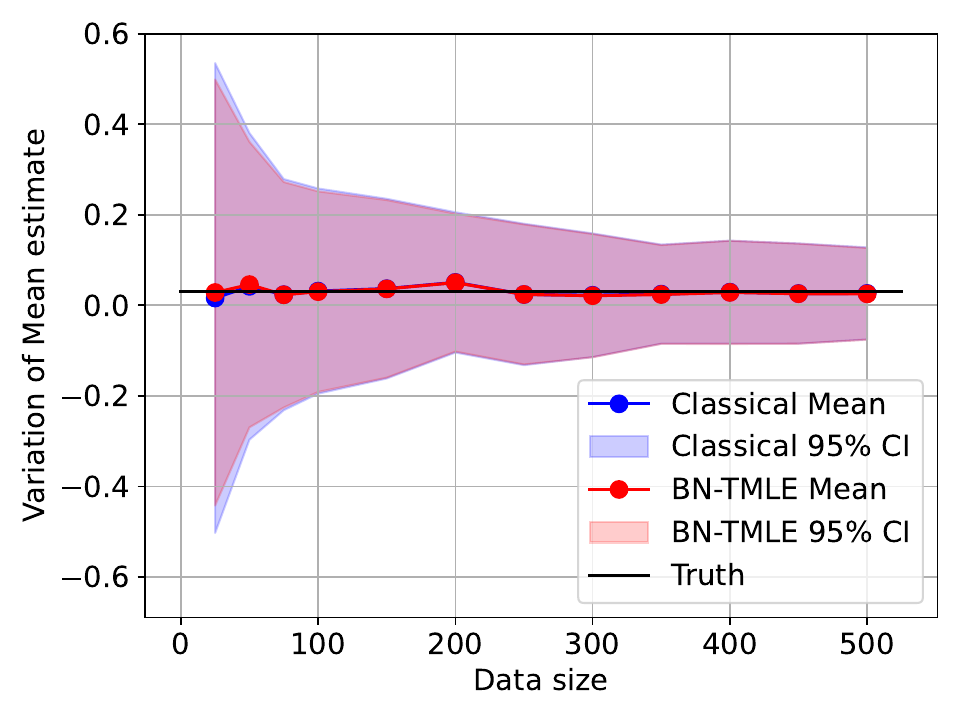}
        
    }
    \subfigure[OMS: Width variation]{
        \includegraphics[width=0.3\textwidth]
        {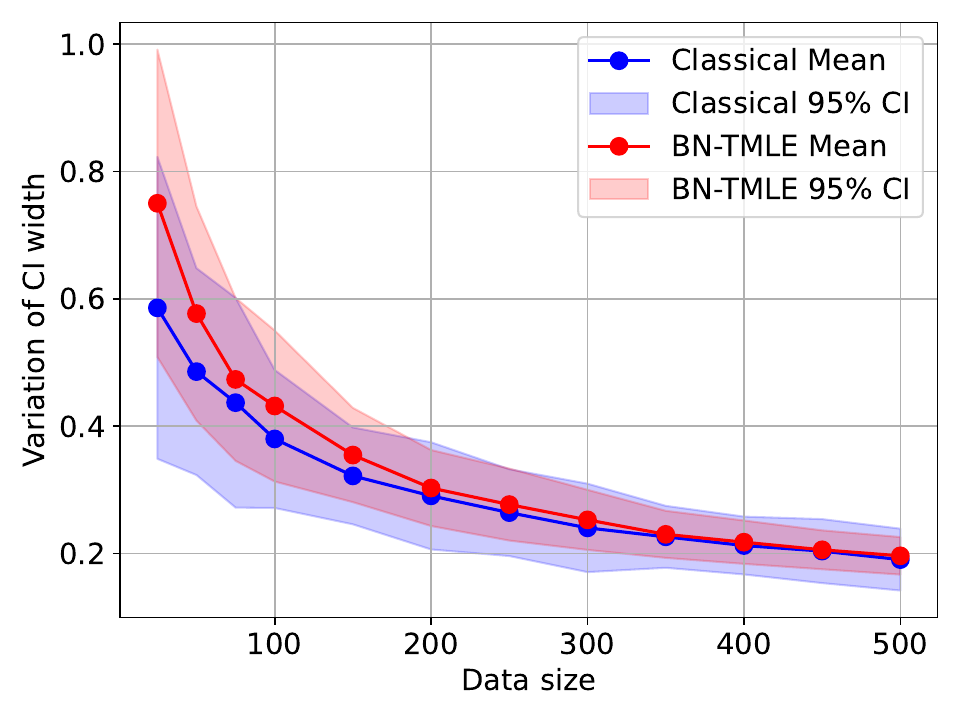}
        
    }
    \subfigure[OMS: Coverage]{
        \includegraphics[width=0.3\textwidth]
        {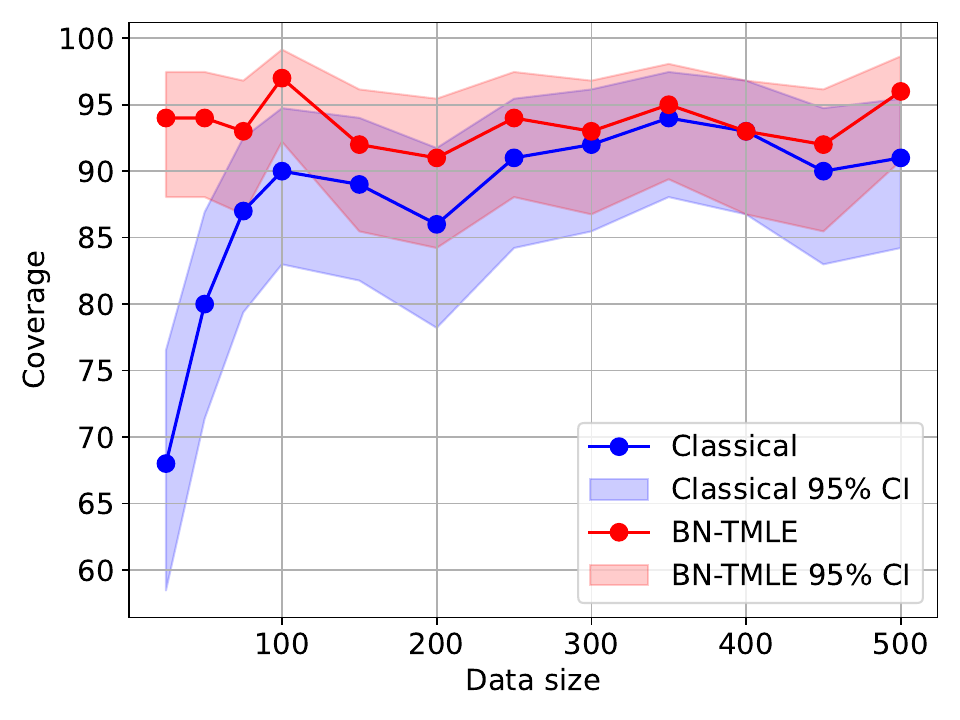}
        
    }
    \vspace{0.5cm}
    \subfigure[OPMS: Mean variation]{
        \includegraphics[width=0.3\textwidth]
        {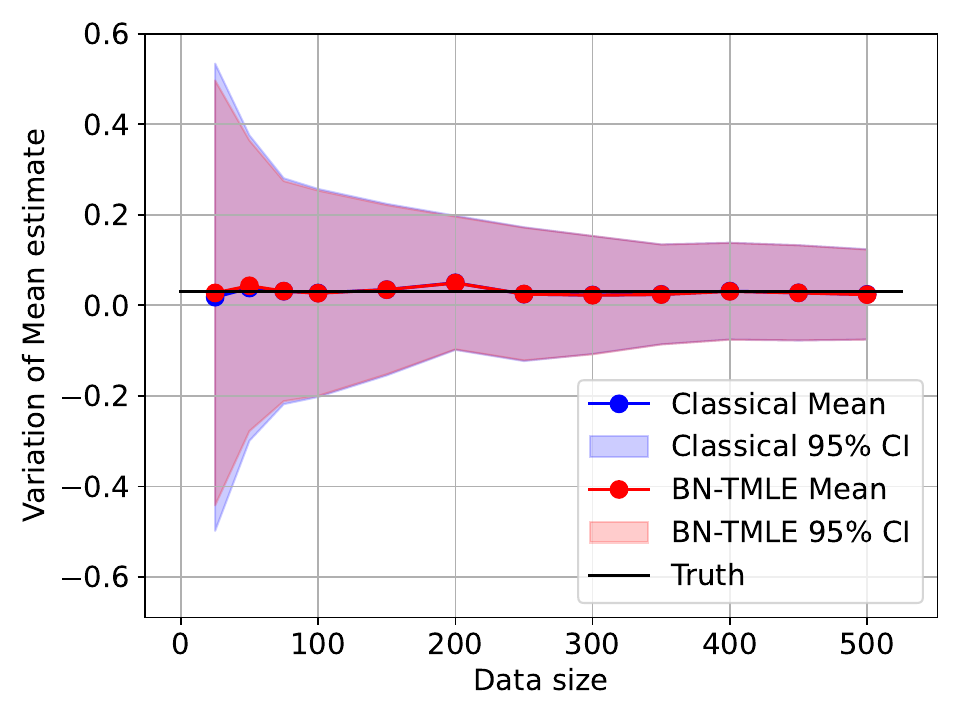}
        
    }
    \subfigure[OPMS: Width variation]{
        \includegraphics[width=0.3\textwidth]
        {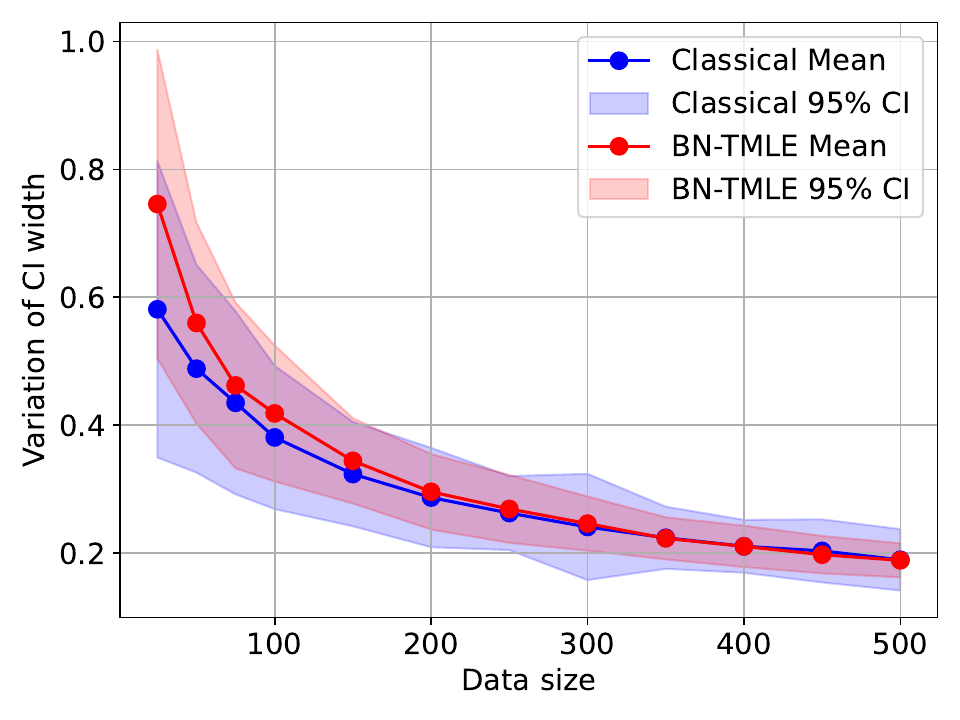}
        
    }
    \subfigure[OPMS: Coverage]{
        \includegraphics[width=0.3\textwidth]
        {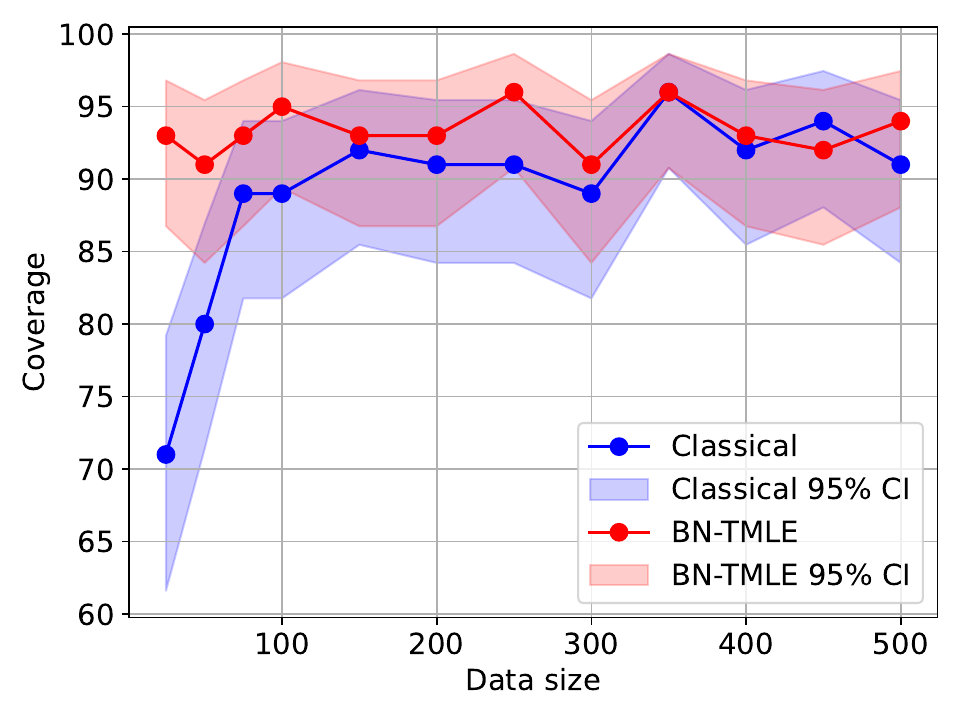}
        
    }
    \caption{Comparison of variation of mean estimate, variation in width of 95\% confidence interval, and coverage percentages of truth across 100 replications of various data sizes with and without considering model discrepancy in outcome and propensity data generation (NMS: No model discrepancy in outcome and propensity models, OMS: Model discrepancy in outcome model only, OPMS: Model discrepancy in both outcome and propensity models), where the True ATE is 0.03 (3\%)}
    \label{fig:datasize_3}
\end{figure}

\begin{figure}[htbp]
    \centering
    \subfigure[NMS: Mean variation]{
        \includegraphics[width=0.3\textwidth]
        {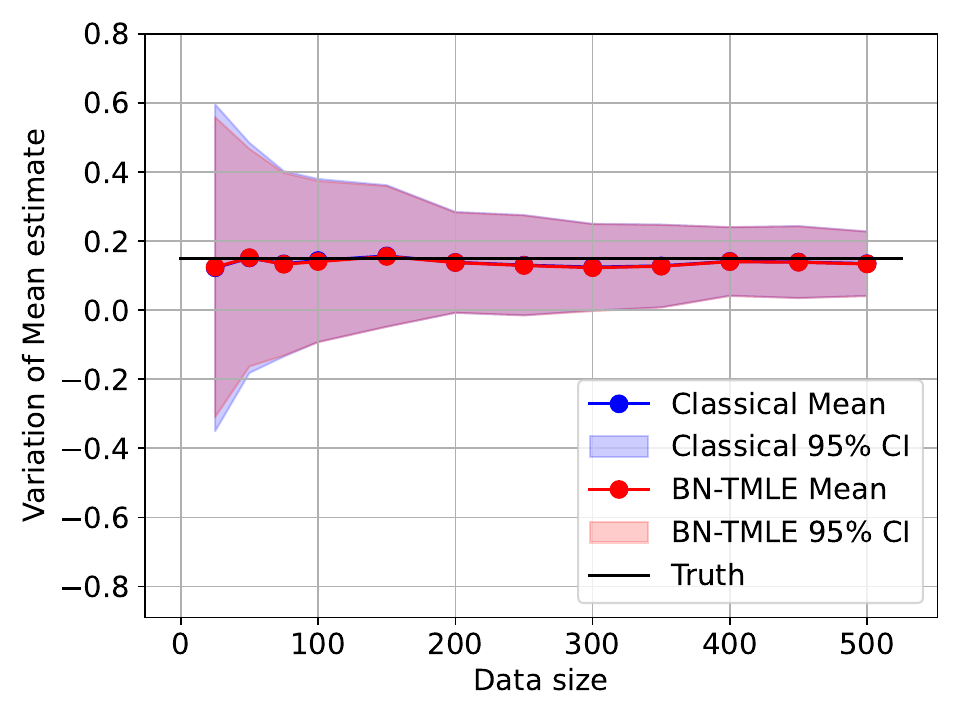}
        
    }
    \subfigure[NMS: Width variation]{
        \includegraphics[width=0.3\textwidth]
        {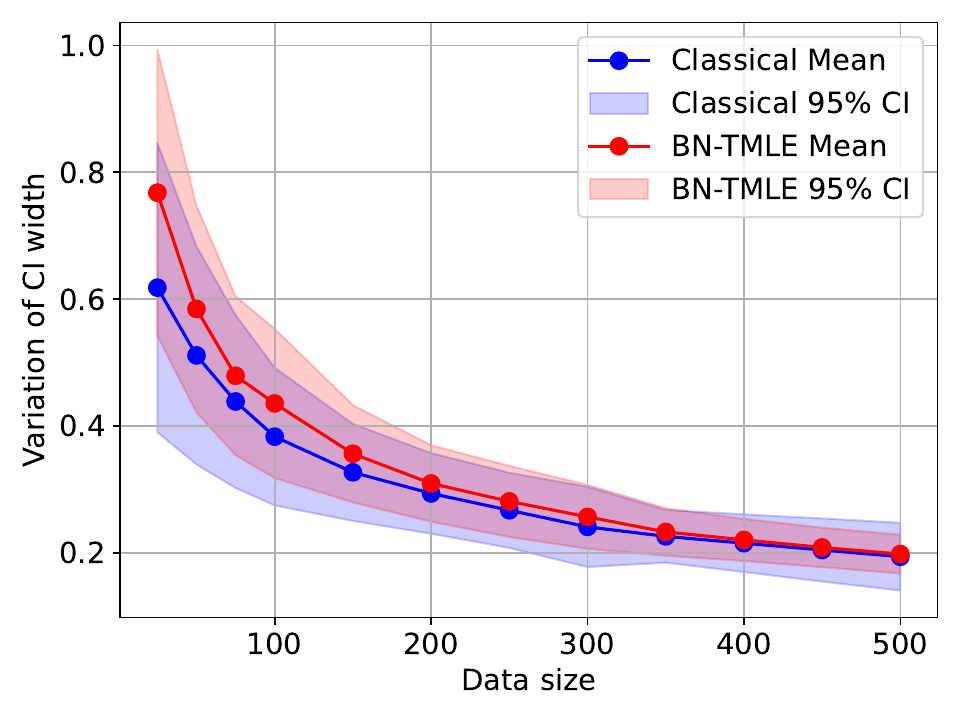}
        
    }
    \subfigure[NMS: Coverage]{
        \includegraphics[width=0.3\textwidth]
        {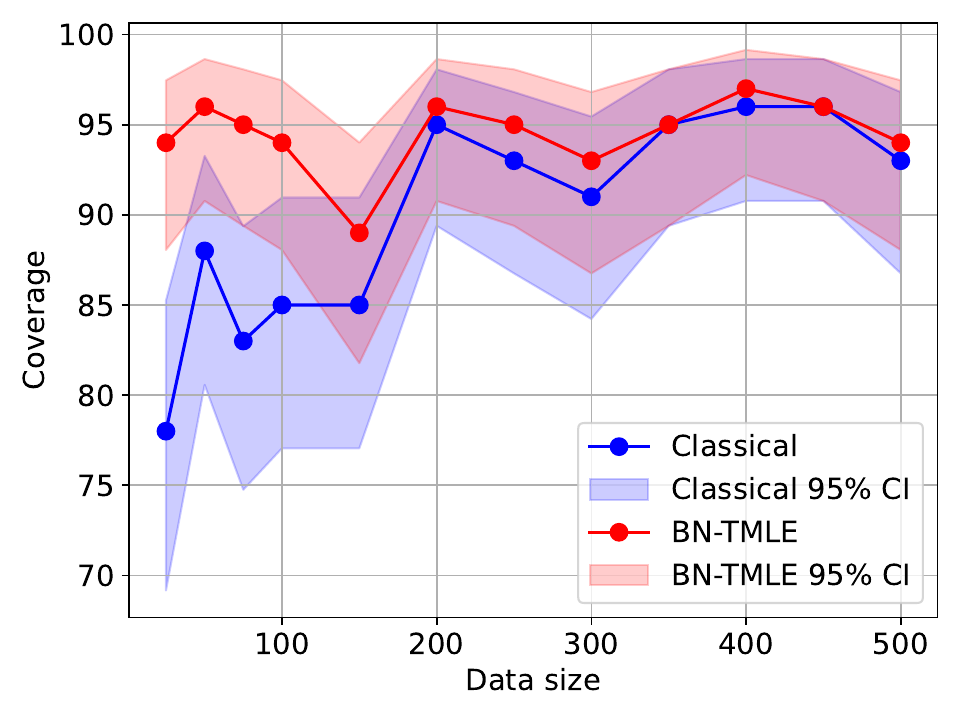}
        
    }
    \vspace{0.5cm}
    \subfigure[OMS: Mean variation]{
        \includegraphics[width=0.3\textwidth]
        {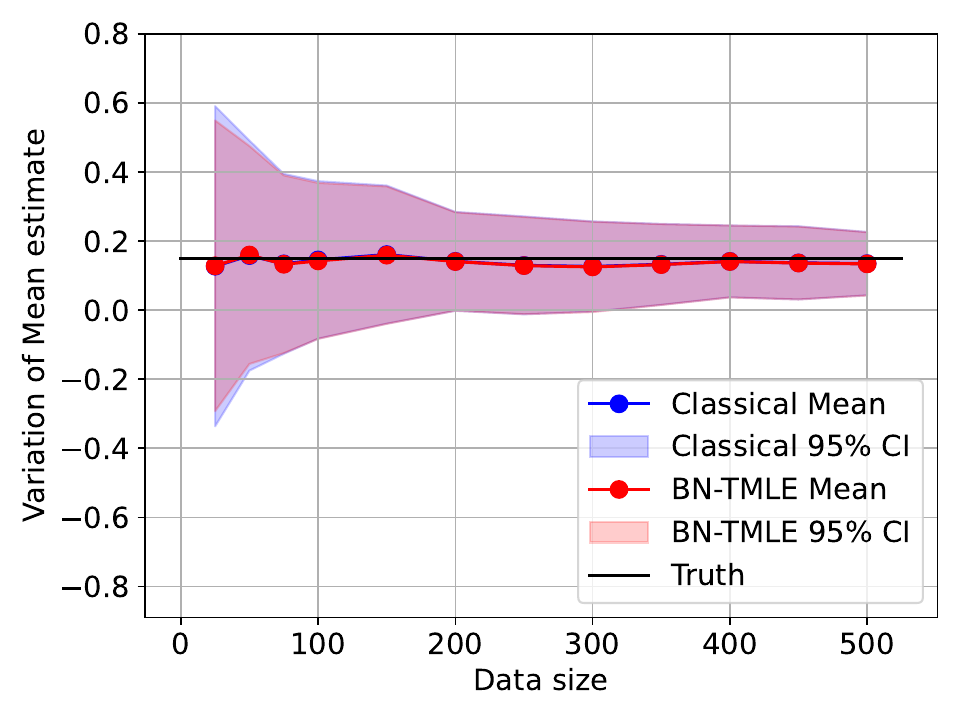}
        
    }
    \subfigure[OMS: Width variation]{
        \includegraphics[width=0.3\textwidth]
        {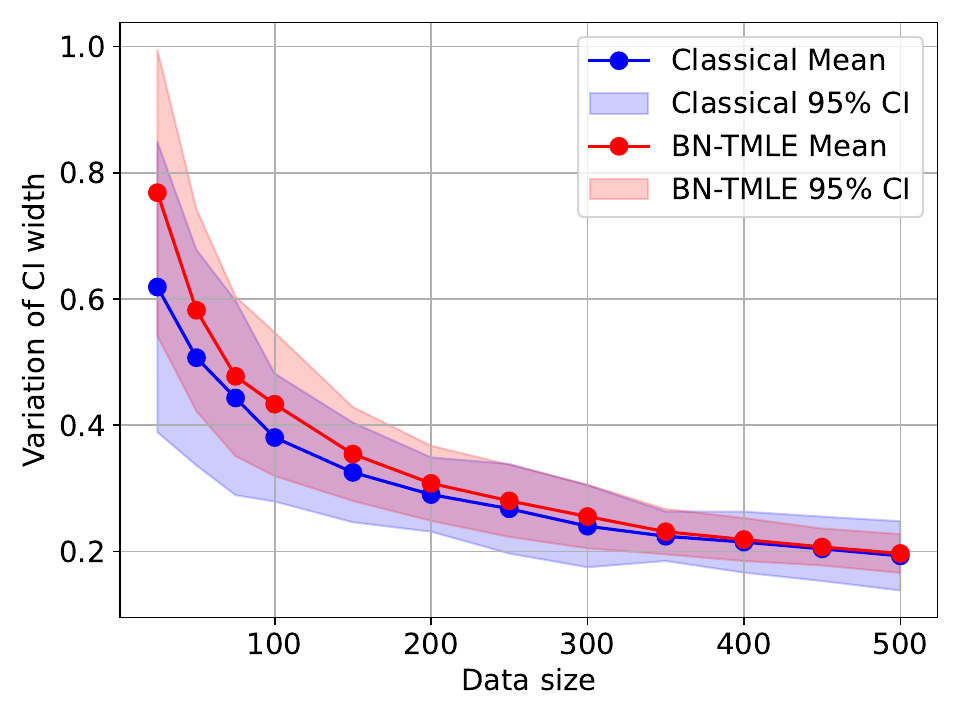}
        
    }
    \subfigure[OMS: Coverage]{
        \includegraphics[width=0.3\textwidth]
        {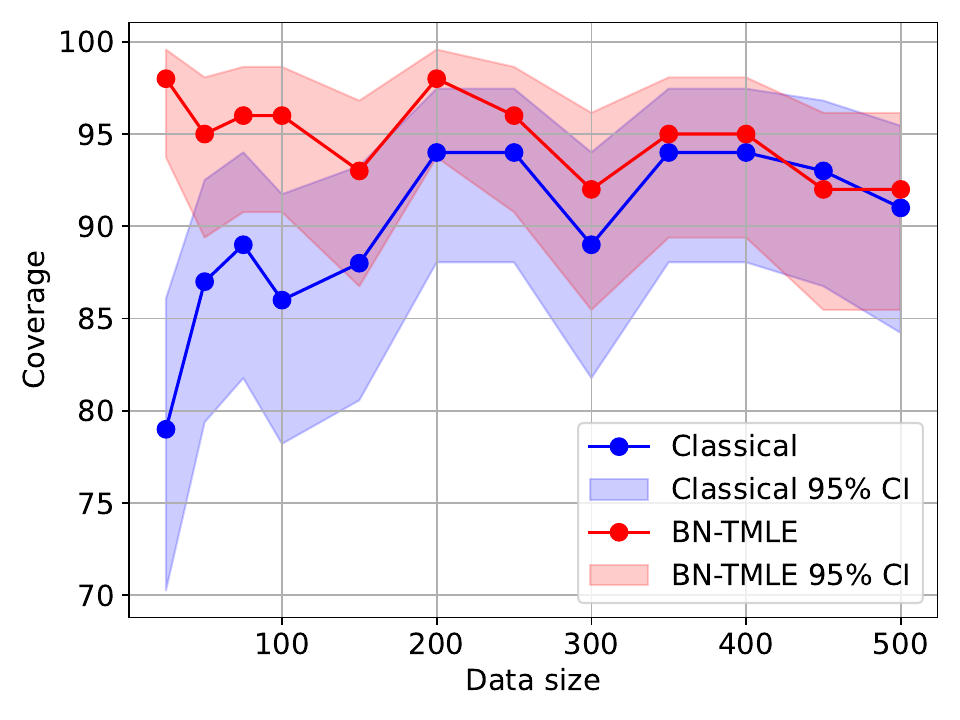}
        
    }
    \vspace{0.5cm}
    \subfigure[OPMS: Mean variation]{
        \includegraphics[width=0.3\textwidth]
        {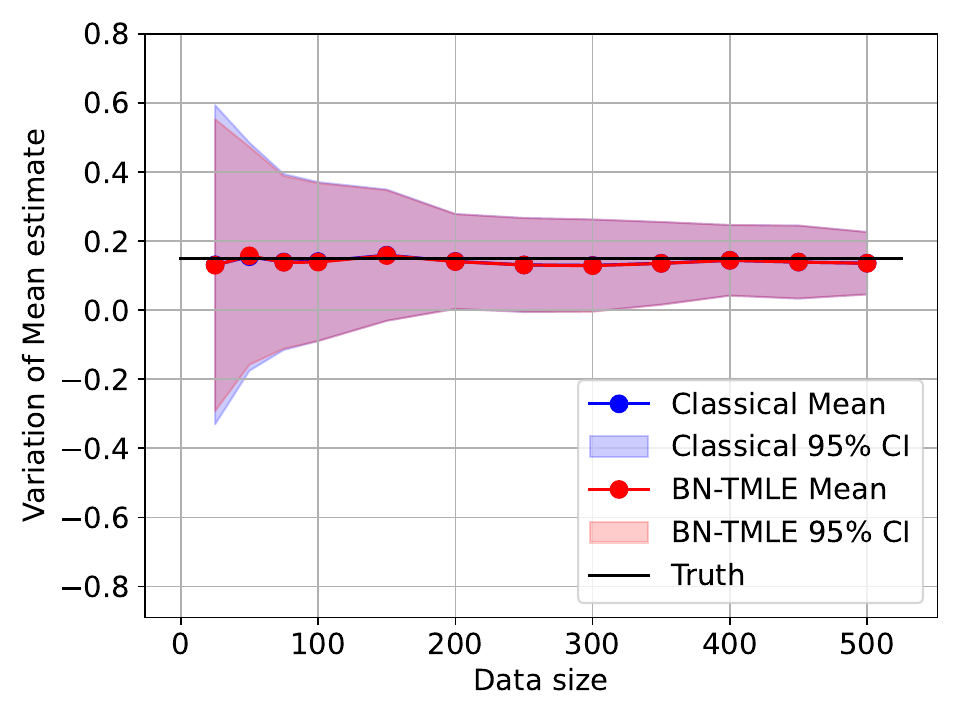}
        
    }
    \subfigure[OPMS: Width variation]{
        \includegraphics[width=0.3\textwidth]
        {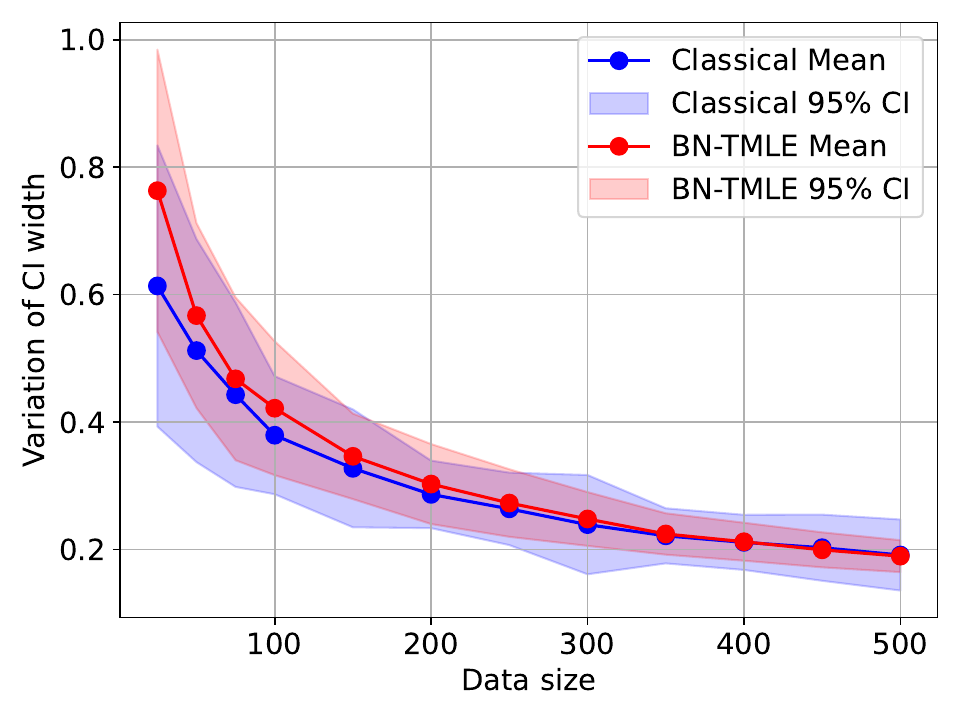}
        
    }
    \subfigure[OPMS: Coverage]{
        \includegraphics[width=0.3\textwidth]
        {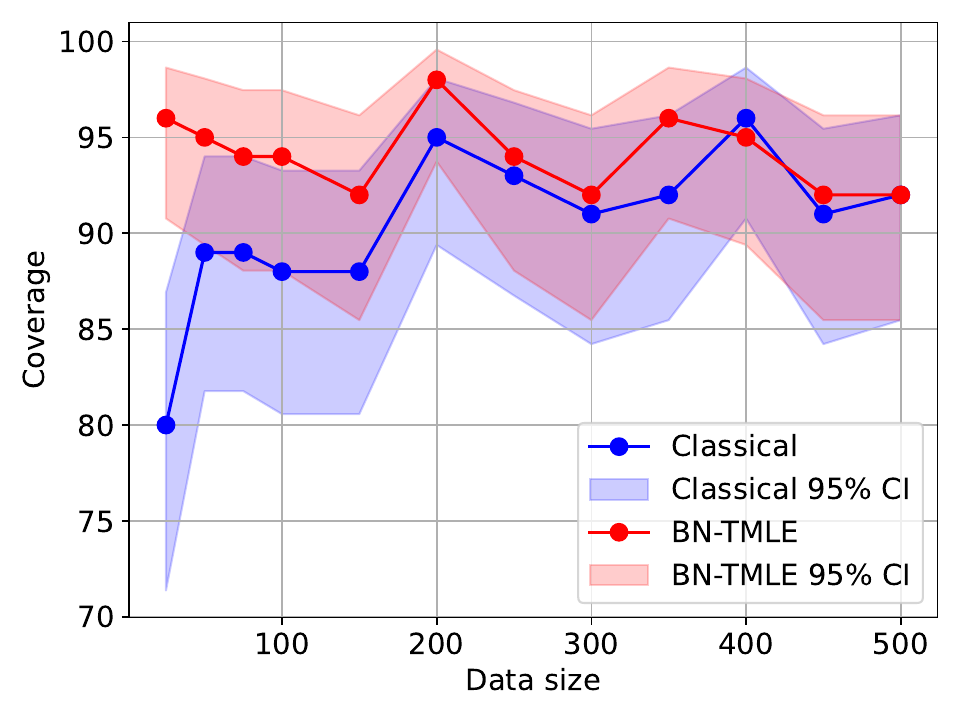}
        
    }
    \caption{Comparison of variation of mean estimate, variation in width of 95\% confidence interval, and coverage percentages of truth across 100 replications of various data sizes with and without considering model discrepancy in outcome and propensity data generation (NMS: No model discrepancy in outcome and propensity models, OMS: Model discrepancy in outcome model only, OPMS: Model discrepancy in both outcome and propensity models), where the True ATE is 0.15 (15\%)}
    \label{fig:datasize_15}
\end{figure}

Therefore, at each level of model misspecification, we provide three visualizations: (1) variation of mean estimate and associated 95\% CI at each data size computed using 100 replications, (2) variation in the width of 95\% CI and associated 95\% CI at each data size computed using 100 replications, and (3) coverage percentage computed using 100 replications at each data size  and associated 95\% CI . We compared the performance of both classical TMLE and BN-TMLE implementations.  Figure \ref{fig:datasize_3}  and \ref{fig:datasize_15} show the effect of data sizes on ATE estimation across three levels of model misspecifications when the true ATE values are 3\% and 15\% respectively.

\textbf{Discussion:} We can make the following observations from Figures \ref{fig:datasize_3} and \ref{fig:datasize_15}. 
\begin{itemize}
    \item When data size is small (25 data points), the 95\% CI of the mean variation from BN-TMLE is slightly narrower when compared to classical TMLE resulting in a slightly more precise estimation of mean ATE value; however, the difference is negligible starting from 50 data points
    \item  At small data sizes,  the average width from BN-TMLE (red dots in width variation plots) is higher when compared to classical TMLE (blue dots). However, as data size increased (more than 300 data points), the average widths of BN-TMLE and classical TMLE converged. This indicates that BN-TMLE at small data sizes, on average, produces wider confidence intervals of ATE when compared to classical implementation
    \item At higher data sizes (more than 200 data points), the 95\% CIs of width associated with BN-TMLE were narrower when compared to classical TMLE leading to lower uncertainty in the width of ATE CI. Therefore, BN-TMLE started with a higher average width and wider 95\% CIs when compared to classical analysis, however, at large data sizes, BN-TMLE achieved narrower 95\% CIs. Since the 95\% CI of widths of classical and BN-TMLE overlap, it is possible that for a random dataset, classical TMLE produce a narrower or a similar CI when compared to BN-TMLE. This scenario was observed in Figure \ref{fig:ate_bin}, where widths from both classical and BN-TMLE approaches are similar.  
    \item BN-TMLE outperformed classical TMLE in coverage, evidently more in small data sizes. For example, at a data size of 25 points, the 95\% CIs of the coverages of classical and Bayesian approaches did not overlap, which indicates that BN-TMLE outperformed classical implementation, and was able to better capture the true ATE value. The mean coverage percentages of BN-TMLE were generally higher than those of classical TMLE at all data sizes; however, the differences between Bayesian and classical TMLE became smaller at larger data sizes.
\end{itemize}

In summary, BN-TMLE and classical analysis produces similar mean estimates at all data sizes except for very small data sizes, where BN-TMLE may produce an estimate that is closer to the truth. At small data sizes, BN-TMLE produces wider CIs leading to a higher chance of capturing the underlying true ATE. At higher data sizes, BN-TMLE is more robust leading to lower uncertainty (narrow 95\% CI) in the ATE CI width. We can conclude from the above study that Bayesian TMLE implementation produces more accurate ATE estimates in small data regimes when coverage is considered as a performance metric and more robust estimates in large data regimes. Since TMLE is applicable to both randomized trials and observational datasets, some practical benefits of BN-TMLE include faster ATE estimations when data generation (especially in randomized trials) is expensive. In observational studies, when sequential data is available, BN-TMLE can be used to arrive at causal conclusions much earlier when compared to classical TMLE. The wider 95\% CI that BN-TMLE produces in small data regimes acts as a safeguard against making an inaccurate causal conclusion in small data regimes.

\subsection{Effect of fluctuation model formulation}
\label{subsec:fluc_exp}

\begin{figure}[!htb]
    \centering
    \subfigure[Binary outcome]{
        \includegraphics[width=0.47\textwidth]
        {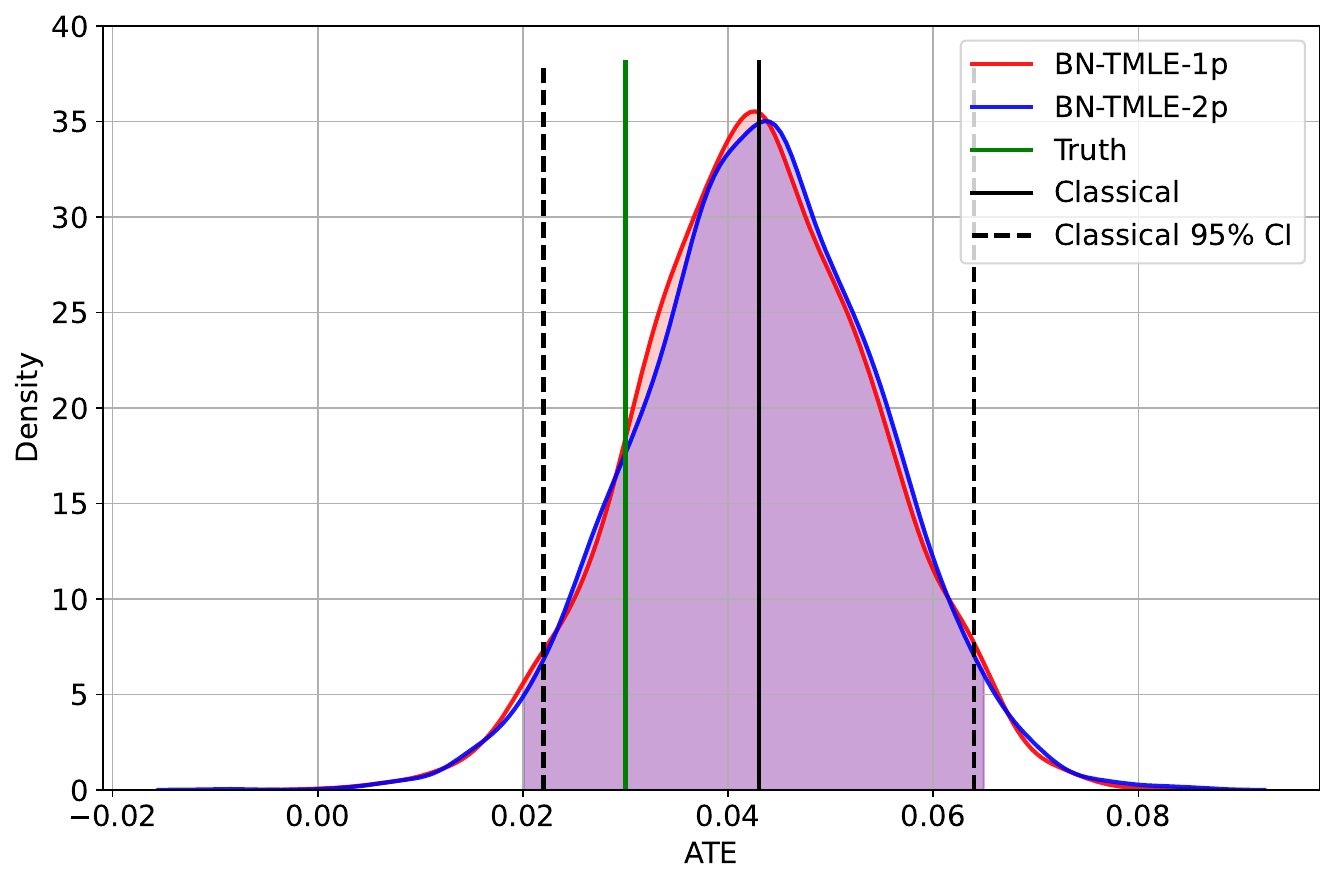}
        
    }
    \subfigure[Continuous outcome]{
        \includegraphics[width=0.47\textwidth]
        {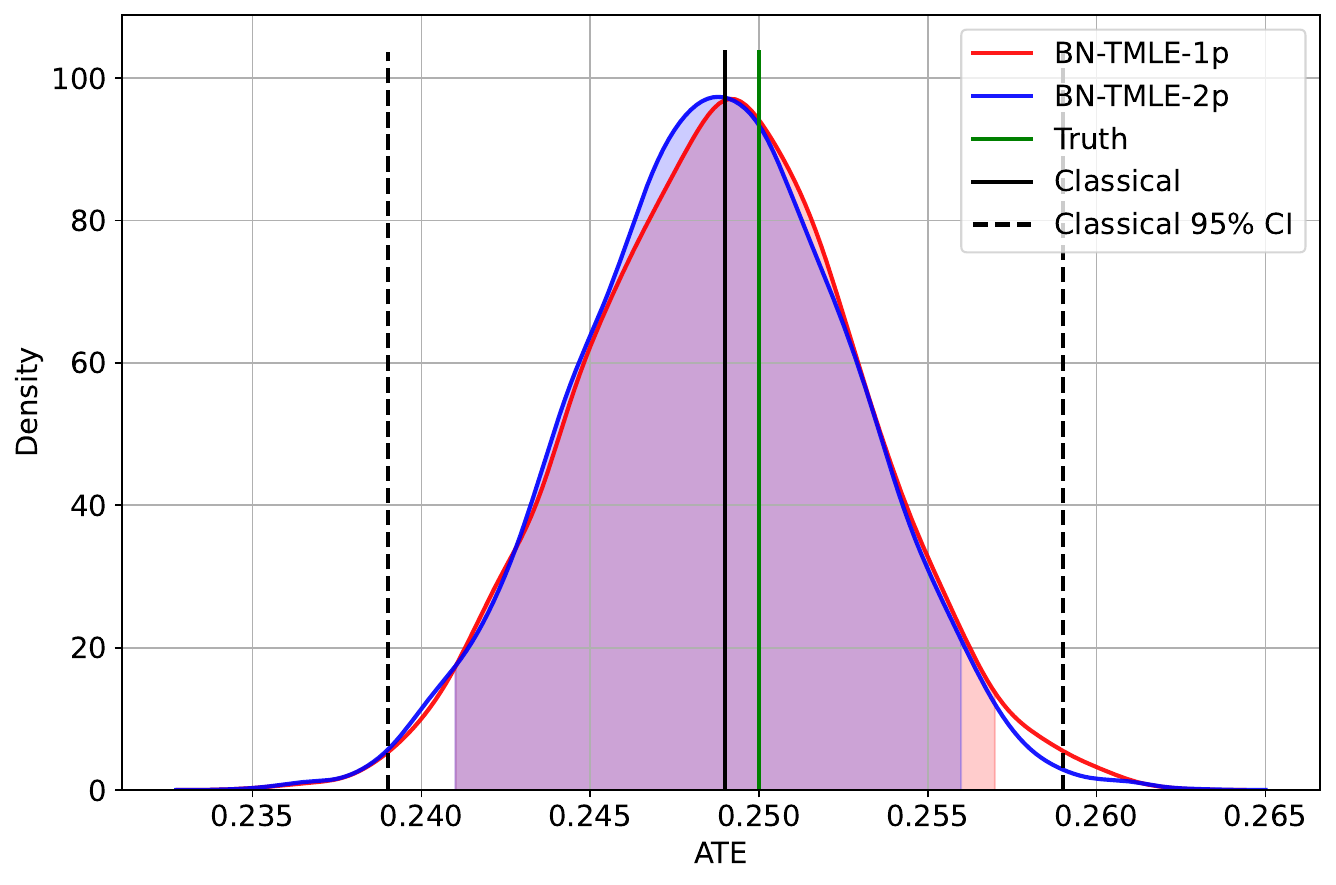}
        
    }


    \caption{Comparison of ATE distributions using the BN-TMLE framework with one and two fluctuation parameter formulations against classical results}
    \label{fig:tmle-compare}
\end{figure}

Section \ref{subsec:bn-tmle} discussed BN-TMLE implementation for both one-parameter and two-parameter fluctuation parameters, referred here as BN-TMLE-1p and BN-TMLE-2p respectively. We should note that the BN-TMLE analysis in Sections \ref{subsec: case} and \ref{subsec:data} used the one-parameter formulation (BN-TMLE-1p). As discussed in Section \ref{subsec:bn-tmle}, BN-TMLE-2p has one additional inference parameter when compared to BN-TMLE-1p (in the fluctuation model formulation). To assess the impact of an additional inference parameter on the ATE distribution, we repeated the Bayesian network TMLE analysis for the three datasets discussed in Section \ref{subsec: case} (binary and continuous outcomes), and the ATE distributions associated with BN-TMLE-1p and BN-TMLE-2p along with the classical TMLE results (both mean and 95\% CI) are available in Figure \ref{fig:tmle-compare}.  We should note that both the one-parameter and two-parameter classical TMLE implementations produced the exact same results.  Table \ref{tab:fluc_12} provides numerical values of mean and 95\% CI of ATE for all three case studies. We can observe that the ATE estimates with BN-TMLE-1p and BN-TMLE-2p are similar for both binary and continuous outcomes. Even though BN-TMLE-2p has an additional parameter, it did not result in any increased variance of the corresponding ATE distribution. Hence, in the case of a binary treatment, either a one or a two parameter fluctuation model formulation can be chosen for uncertainty quantification in causal effect estimation.

\begin{table}[htbp]
  \centering
  \caption{Comparison of ATE estimates with one and two-parameter fluctuation models}
  \vspace{0.3cm}
  \scalebox{0.68}{
    \begin{tabular}{l|c|c|c|c}
    \toprule
    \textbf{Outcome} & \textbf{Truth} & \textbf{Classical} & \textbf{BN-TMLE-1p} & \textbf{BN-TMLE-2p} \\
    \midrule
    \textbf{Binary} & 0.03 & 0.043 [0.022, 0.064] & 0.043 [0.02, 0.065] & 0.043 [0.02, 0.065] \\
    \textbf{Continuous} & 0.25 & 0.249 [0.239, 0.259] & 0.249 [0.241, 0.257] & 0.249 [0.241, 0.256] \\
    \bottomrule
    \end{tabular}}
  \label{tab:fluc_12}
\end{table}

\section{Conclusion}
\label{sec:conc}

This paper discussed Bayesian implementation of the widely popular Targeted Maximum Likelihood Estimation (TMLE) algorithm for uncertainty quantification of the causal effect using a sample-based probability distribution as opposed to statistical confidence intervals. The training of the fluctuation model (the ``targeting" step in TMLE) produced a unique set of challenges for Bayesian inference as the output(true outcome) data are point values while the inputs (initial outcome predictions and clever covariates) are distributions obtained from the outcome and propensity models. We proposed two methods (B-TMLE-M and B-TMLE-SS)  to overcome these challenges where we used only the mean estimates, and summary statistics (such as mean and standard deviation) of the initial outcome predictions and clever covariate values. In addition, we proposed a Bayesian network approach for TMLE implementation (BN-TMLE) where all the three TMLE models are combined and trained simultaneously.  We compared the performance of these three Bayesian TMLE approaches against two simulation datasets with both binary and continuous outcomes against the classical TMLE results and true valies. The three Bayesian TMLE approaches were in good agreement with classical implementation, and among the three Bayesian approaches, BN-TMLE produced optimum results with a lower variance in the ATE distribution when compared to the other two approaches as BN-TMLE considered the full probability distributions of initial outcome and clever covariates while B-TMLE-M and B-TMLE-SS considered only the mean and summary statistics of predictions respectively. 

In addition, we conducted  a simulation analysis to investigate the effect of data size and model misspecifications on the performance of BN-TMLE and classical TMLE, and results showed that BN-TMLE outperforms classical TMLE in small data regimes using coverage as a performance metric, and in large data regimes, BN-TMLE produced more robust ATE confidence intervals when compared to classical implementations. However, both the Bayesian TMLE and classical TMLE approaches produced similar mean ATE estimates. Also, we investigated the performance of one-parameter and two-parameter fluctuation model formulations in BN-TMLE on ATE distributions, and results concluded that both the formulations produced similar ATE distributions. 

Future work should also extend the proposed Bayesian TMLE approaches to other data types of treatment and outcome variables such as categorical/continuous treatments and a categorical outcome, and other variants of TMLE such as Longitudinal TMLE and Survival TMLE for causal inference with longitudinal data and time-to-event data respectively. Moreover, future work should also investigate efficient Bayesian inference techniques for large-scale data processing such as mini-batching (\cite{sekkat2021quantifying}) and early stopping in MCMC-based inference for scalable Bayesian TMLE implementation.

\section*{Acknowledgement}
The authors thank Nicholas Tomasino for valuable discussions during the early stages of this work, and Jagadeesh Macherla at App Orchid for providing resources and offering support towards this work.

\section*{Funding sources}
This research did not receive any specific grant from funding agencies in the public, commercial, or not-for-profit sectors.


\bibliographystyle{apalike} 
\bibliography{references}

\newpage
\appendix
\section{Introduction to Bayesian networks}
\label{app:bn}

A Bayesian Network (BN) is a directed acyclic graphical model that represents dependence between a set of variables (typically represented through a joint probability distribution) using a combination of marginal and conditional probability distributions \citep{koller2009probabilistic}. The variables are represented as nodes in a Bayesian networks and the edges between them represent statistical dependence quantified through conditional probability distributions. Consider $n$ variables represented as $\mathbf{X}=\{X_1,X_2,...,X_n\}$. These $n$ variables are represented as $n$ nodes in a Bayesian network. Dependence between two variables $X_i$ and $X_j$ is represented using a directed edge from $X_i$ to $X_j$. $X_i$ is referred to as a parent node of $X_j$ while $X_j$ is referred to as a child node of $X_i$. Using the Bayesian network, the joint probability distribution over $\mathbb{X}$ can be written as shown in Equation \ref{eqn:bn}.

\begin{equation}
\label{eqn:bn}
    p(X_1,X_2,...,X_n)= \prod_{i=1}^{n} p(X_i| \Pi_{X_i})
\end{equation}

In Equation \ref{eqn:bn}, $\Pi_{X_i}$ represents the set of all parent nodes of $X_i$. For a node without any parent nodes, the conditional distribution is replaced with an unconditional(or marginal) distribution, i.e., $p(X_i| \Pi_{X_i}) = p(X_i)$.  

The directional dependence between nodes in a Bayesian network results in a hierarchy of variables. Sometimes, terms such as upstream and downstream variables are used to represent the location of a variable in the hierarchy with respect to another variable in the Bayesian network \citep{zou2009granger}. An upstream variable of a variable $X_i$ is any variable that is a parent node of $X_i$ or an ancestor node (e.g., parent node of a parent node). Similarly, a downstream variable of $X_i$ is any variable that is a child node of $X_i$ or any descendant node (e.g., child node of a child node).  Given data on one or a few variables (called observed variables), the BN can be used to infer distributions of unobserved variables conditioned on the observed variables at the available data using Bayesian inference. The exact and approximate inference algorithms discussed in Section \ref{subsec:bayes} can be used for inferring unobserved variables using data on observed variables. 

Typically, the dependence relationships between nodes in a Bayesian network are stochastic, and they are represented through conditional probability distributions, i.e., probability distributions of child nodes conditioned on the values of their parent nodes. Sometimes, nodes with deterministic relationships with the parent nodes (typically represented using mathematical equations) are also shown in a Bayesian network to provide a graphical representation of dependence between variables. Such nodes with deterministic relationships with parent nodes are referred to as deterministic nodes \citep{geiger1990identifying}. 

\section{Learning Bayesian networks}
\label{app:learn_bn}

Primarily, there are three approaches for constructing a Bayesian network given a set of variables, any available data and/or dependence relationships (such as mathematical models and expert knowledge) between them: (1) using dependence relationships, (2) using data, and (3) using a combination of dependence relationships and data \citep{kitson2023survey}.  In the first approach, we categorize available dependence relationships into two categories - qualitative and quantitative. Quantitative relationships provide mathematical relationships between variables (such as the outcome model discussed in Equation \ref{eqn:outcome}) whereas qualitative relationships provide the direction of dependence without any mathematical relationships between variables. For example, there exists edges from confounders to the treatment and outcome variables, and from the treatment to the outcome variable.  When quantitative relationships are available between variables, the variables become nodes in the Bayesian network, and edges are drawn from the independent variables (such as confounders, treatment variable, and model parameters in Equation \ref{eqn:outcome}) to the dependent variable (outcome variable), and the available mathematical relationships can be used as conditional probability distributions. 

When qualitative relationships are available between variables (direction of edges), nodes and edges can be drawn from these known relationships; however the conditional distributions (specifically their distribution parameters) need to be estimated from data.  The distribution parameters of the conditional distributions are usually modeled as functions of the parent nodes to capture the dependence between a node and its parent nodes. When a functional form of dependence is known (e.g., linear dependence), we assume that functional form and estimate the parameters associated with the functional form. For illustration, consider a Bayesian network with three nodes: $X_1, X_2$ and $X_3$ and two known qualitative relationships, $X_1 \rightarrow X_3$ and $X_2 \rightarrow X_3$. Here, $X_1$ and $X_2$ are parent nodes of $X_3$ and the conditional distribution, $p(X_3|X_1, X_2)$ will need to be estimated from data. If we model the conditional distribution as Normal distribution where the mean is dependent on the parent nodes and an unknown standard deviation as $p(X_3|X_1, X_2) \sim \mbox{Normal}(a + b_1X_1 +b_2X_2, \sigma)$ , then the problem of estimation of conditional distribution simplifies to a parameter estimation problem where the parameters $\{a, b_1, b_2, \sigma\}$ will be estimated from data. This linear dependence is shown for illustration only and we can assume any linear or non-linear dependence (such as a deep neural network) and estimate relevant parameters from data.

When dependence relationships are unavailable but data on variables are available, Bayesian network  structure learning (BNSL) algorithms can be used to identify a Bayesian network (edges between nodes and associated conditional probability distributions) from data.  BNSL is a well-studied topic in the literature, and a detailed review of available algorithms are available in \citet{scanagatta2019survey} and \citet{kitson2023survey}. The hybrid approach is typically used in complex systems where dependence relationships (qualitative and/or quantitative) may exist between a subset of variables, and data are available on the remaining variables. In such a scenario, we construct the overall Bayesian network in two stages. In the first stage, we add edges between nodes derived from available dependence relationships resulting in a partial Bayesian network, and in the second stage, we learn remaining edges between other variables from data using Bayesian network learning algorithms \citep{kitson2023survey}. 

\end{document}